\documentclass[10pt]{article}
\usepackage{amsthm}

\usepackage{graphicx} % Required for inserting images
\usepackage{amsmath}
\usepackage{amssymb}
\usepackage{cite}
\usepackage{booktabs}
\usepackage{amsfonts}
\usepackage{mathrsfs}
\usepackage[latin1]{inputenc}
\usepackage{float}
\usepackage{natbib}
\usepackage{amsmath,amssymb}
\usepackage{latexsym}
\usepackage{epstopdf}
\usepackage{geometry}  
\usepackage{tabularx}
\usepackage[active]{srcltx}
\usepackage{graphicx}
\usepackage{epstopdf}
\usepackage{booktabs}
\usepackage{multirow}
\usepackage{siunitx}
\usepackage{mathtools}
\usepackage{setspace}
\usepackage[
bookmarks=true,         %generates bookmarks for all entries in the "Table of Contents"
bookmarksnumbered=true, %includes chapter-/header-/section-/subsection-/... -
colorlinks=true, pdfstartview=FitV, linkcolor=blue, citecolor=blue,
urlcolor=blue]{hyperref}

\providecommand{\U}[1]{\protect\rule{.1in}{.1in}}

\begin{document}

%\maketitle
\title{\Large Data-Adaptive Multi-Bin Rarefying for Alpha Diversity Association Analysis under Heterogeneous Library Sizes}
\vspace{0.5cm}
\author{Mo Li\thanks{CONTACT: Mo Li;  Email: mo.li@louisiana.edu}  }
\date{%
Department of Mathematics, University of Louisiana at Lafayette, Lafayette, LA 70504, USA\\
% [2ex]%
% \today
}

\maketitle

\begin{abstract}
Alpha diversity association analysis is sensitive to heterogeneous library sizes, and overall rarefying to a common depth can either discard many reads from high-depth samples or exclude low-depth samples. We propose \texttt{MBRarefy}, a data-adaptive framework for selecting multi-bin rarefying designs. The method chooses library size cutpoints by minimizing an objective that combines residual within-bin library size dependence with a fragmentation penalty, subject to a minimum bin size constraint. The framework supports both fixed-\(K\) and varying-\(K\) cutpoint selection. In T-cell receptor (TCR) motivated simulations, the primary fixed-\(K\) procedure controlled depth-induced false positives and improved power relative to overall rarefying when group and depth confounding was absent. Applications to TCR immunosequencing and wild baboon gut microbiome data illustrate its ability to reduce residual library size dependence while improving read and sample retention.
\end{abstract}

\noindent

%\maketitle

\section{Introduction}\label{sec:intro}

% alpha diversity association + heterogeneous library sizes
Alpha diversity is widely used to summarize within-sample richness and evenness in ecology, microbiome sequencing, T-cell/B-cell receptor repertoire sequencing, and other high-throughput count-profile studies. These summaries are often tested for association with clinical, host, or environmental covariates \citep{willis2019rarefaction, kers2022power, mika2025comprehensive}. However, estimated alpha diversity is sensitive to sampling effort. In sequencing studies, per-sample total counts, often referred to as sequencing depth or library size, can vary substantially because of technical and experimental factors. Samples with larger library sizes are generally capable of revealing more different features and can therefore appear more diverse even when underlying biological diversity is unchanged \citep{roswell2021conceptual}. This association becomes particularly problematic when library size is also associated with the phenotype or exposure of interest. An apparent association between the phenotype and alpha diversity may arise partly from their common relationship with sequencing depth rather than from a genuine biological effect, reducing interpretability and reproducibility \citep{willis2019rarefaction}.

% why single-depth rarefying is not enough
% A wide range of library size normalization strategies has therefore been proposed. Scaling and compositional approaches are commonly discussed for relative abundance analyses \citep{gloor2017microbiome, weiss2017normalization, mcknight2019methods}, while rarefying remains widely used for alpha diversity estimation. Because it produces diversity summaries on a comparable sampling scale and aligns with classical ecological sampling theory \citep{hurlbert1971nonconcept, willis2019rarefaction}. 
% Rarefying addresses heterogeneity in sampling depth by subsampling reads from the observed library, but it cannot correct biases that arise during library preparation or sequencing.
% In practice, sequencing reads are not necessarily an unbiased sample of the underlying molecular population: sequence composition can affect amplification and sequencing efficiency, with GC-rich and AT-rich fragments in particular showing systematic under-representation in high-throughput sequencing data \citep{aird2011analyzing, benjamini2012summarizing, ross2013characterizing}. Consequently, equalizing the number of observed reads across samples does not remove feature-specific representation biases already present in the sequenced libraries. 
A wide range of library size normalization strategies has therefore been proposed. Scaling and compositional approaches are commonly discussed for relative-abundance analyses \citep{gloor2017microbiome, weiss2017normalization, mcknight2019methods}, while rarefying remains widely used for alpha diversity estimation because it places diversity summaries on a comparable sampling scale and aligns with classical ecological sampling theory \citep{hurlbert1971nonconcept, willis2019rarefaction}. Rarefying addresses heterogeneity in observed sampling depth, but it cannot correct feature-specific representation biases introduced during library preparation or sequencing \citep{aird2011analyzing, benjamini2012summarizing, ross2013characterizing}.

Even when the primary concern is library size heterogeneity rather than feature-specific sequencing bias, rarefying typically relies on a single global depth for all samples. This creates a trade-off between retaining low-depth samples and preserving information in samples with larger libraries. A lower rarefying depth retains more samples, but at the cost of discarding a larger fraction of reads from samples with large library sizes. A higher rarefying depth reduces this within-sample loss among retained samples, but removes samples whose library sizes are below the selected threshold. The rarefying depth is often selected heuristically, and different choices can yield different conclusions. Moreover, alpha diversity estimates obtained after rarefying can remain associated with the samples' original library sizes, with several studies reporting positive associations between library sizes and alpha diversity computed from rarefied data \citep{lang2018t, willis2019rarefaction, li2024multi}. Thus, single-depth overall rarefying may reduce variation in nominal sampling effort without fully eliminating residual depth dependence.

% what we have is multi-bin, what can it do and why it is still not enough
The prior work from \citet{li2024multi} introduced a multi-bin rarefying approach for alpha diversity association analysis under heterogeneous library sizes. Samples are stratified by library size, rarefied within each bin to a bin-specific depth, and the resulting bin-specific association estimates are combined via meta-analysis.
By analyzing samples at a common rarefying depth within each bin, the approach is intended to substantially reduce the residual association between alpha diversity estimates and library size before combining the stratum-specific effects. It can also preserve substantially more information for moderate and high depth samples than single-depth rarefying while retaining lower-depth samples in the analysis.
However, the library size cutpoints must be specified in advance. Because these cutpoints determine bin sizes, bin-specific rarefying depths, and residual depth dependence, their selection is a statistical design problem rather than a purely operational choice. The original approach provided practical guidance for choosing cutpoints, but the selection remained heuristic and user-specified, without an explicit criterion for comparing candidate partitions or an automated procedure that adapts to the observed data.

% methodological gap that the cut points choice is important
% The selection of library size cutpoints can therefore be viewed as a statistical design problem rather than a purely operational choice. Although the original multi-bin rarefying approach \citep{li2024multi} provided practical guidance for choosing cutpoints, those recommendations were heuristic and required user judgment. They did not define an explicit criterion for comparing candidate partitions or provide an automated procedure that adapts to the observed data.
% our contribution
To address this gap, we develop \texttt{MBRarefy}, a data-adaptive framework for alpha diversity association analysis under heterogeneous library sizes. We formulate multi-bin cutpoint selection as an optimization problem over a candidate rarefying-depth grid. For each candidate partition, residual within-bin library size dependence is quantified using a sample size weighted average of bin-specific partial coefficients of determination, while a fragmentation penalty discourages excessive or highly unbalanced partitioning. These components define a unified objective that is minimized subject to a hard minimum bin-size constraint. The primary implementation uses a fixed-\(K\) formulation, in which the number of interior cutpoints is specified by the user and their locations are optimized. An optional varying-\(K\) formulation jointly searches over the number and locations of cutpoints. The ordered cutpoint optimization is carried out using \texttt{GAReg} \citep{li2026GAReg}, with \texttt{changepointGA} as the computational backend \citep{li2026changepointga}. The methodology is implemented in the \texttt{MBRarefy} R package, which also supports repeated rarefying \citep{cameron2021enhancing, hu2022rarefaction}, bin-wise association testing, cross-bin meta-analysis, and reproducible analysis of T-cell receptor (TCR), microbiome, and other count profile data.

% numerical evaluation and applications
We evaluate the proposed framework through simulation studies designed to distinguish depth-induced artifacts from genuine biological association. The simulations examine type I error under the null settings, power under alternatives with and without library size dependence, structural recovery, sensitivity to the candidate rarefying-depth grid, stability with respect to stochastic optimization, and robustness across observed richness and Shannon diversity. We further evaluate the method in two real applications from distinct sequencing domains, a TCR immunosequencing study with cytomegalovirus serostatus (CMV) and a wild baboon gut microbiome study. Together, these analyses assess the operating characteristics, robustness, and practical utility of data-adaptive multi-bin rarefying relative to single-depth overall rarefying.

% paper organization
The remainder of the paper is organized as follows. Section~\ref{sec:method} introduces the data-adaptive multi-bin rarefying framework, including the residual library size dependence criterion, fragmentation penalty, and fixed- and varying-\(K\) cutpoint selection procedures. Section~\ref{sec:sim} evaluates the proposed method through simulation studies examining operating characteristics, structural recovery, and sensitivity and stability properties. Section~\ref{sec:cases} presents applications to TCR sequencing and wild baboon gut microbiome data. Section~\ref{sec:Discussion} concludes with a discussion of the main findings, practical considerations, and limitations.

\section{Data-Adaptive Multi-bin rarefying}\label{sec:method}

\begin{figure*}[!h]
  \centering
  \includegraphics[scale=0.44]{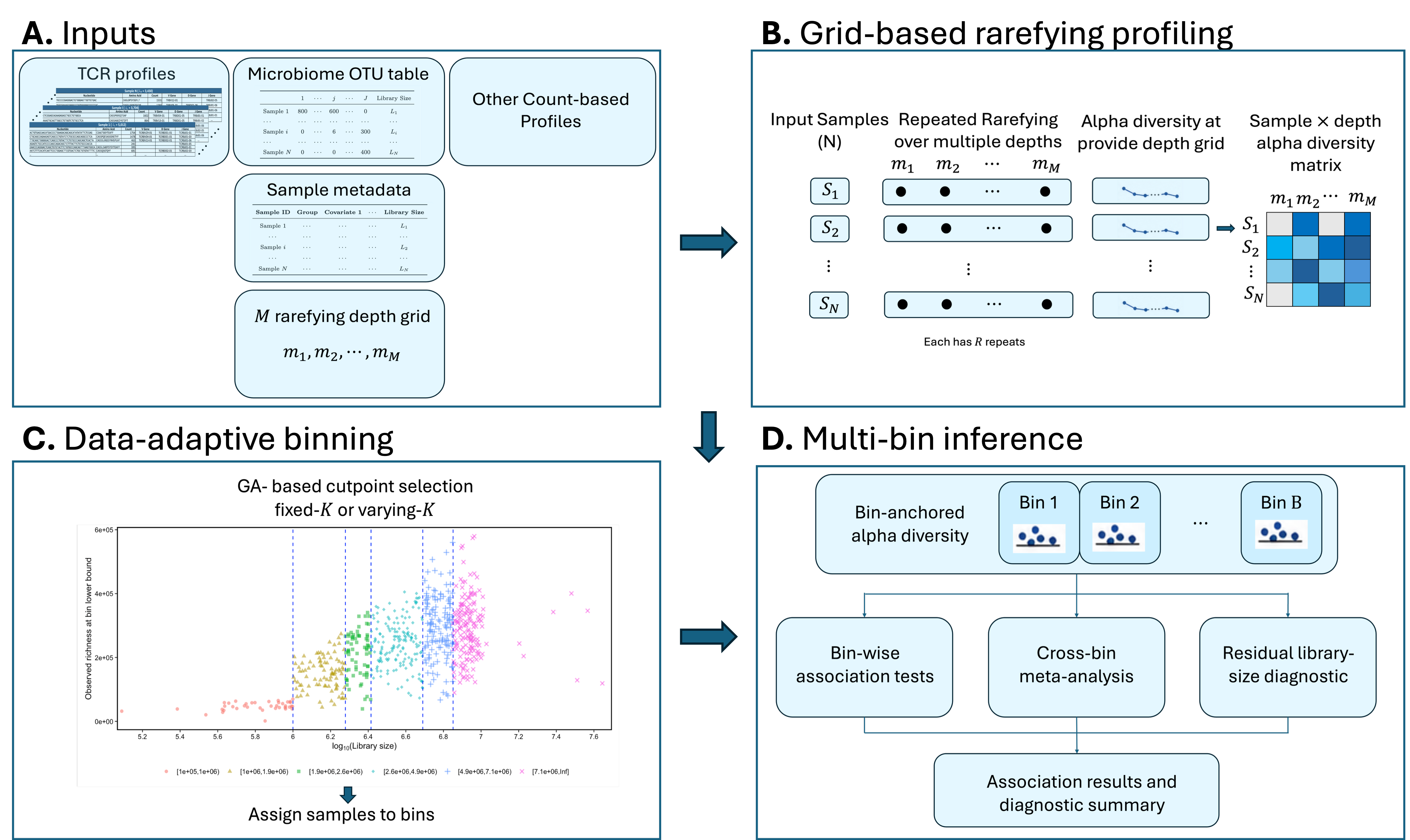}
  \caption{\textbf{Overview of the data-adaptive multi-bin rarefying framework.}
  (A) Feature-count profiles and sample metadata are aligned with a candidate rarefying depth grid.
  (B) Repeated rarefying across candidate depths produces sample-by-depth alpha diversity profiles.
  (C) Fixed-\(K\) or varying-\(K\) optimization selects library size cutpoints by minimizing an objective that combines residual depth dependence and a sample-fragmentation penalty.
  (D) The selected bin-anchored alpha diversity values are used for within-bin association testing, cross-bin meta-analysis, and residual library size diagnostics.}
  \label{fig:mbr_workflow}
\end{figure*}

Figure~\ref{fig:mbr_workflow} summarizes the proposed data-adaptive multi-bin rarefying framework. The procedure first constructs rarefying profiles over a candidate depth grid, then selects library size cutpoints by balancing residual within-bin depth dependence against sample fragmentation. The selected cutpoints define bin-anchored alpha diversity values that are used for bin-wise association testing, cross-bin meta-analysis, and residual library size diagnostics.

\subsection{Rarefying Profiles and Bin-Anchored Diversity}

We consider a sequencing study with \(N\) samples measured over \(p\) features, such as sequence variants, taxa, genes, or immune-receptor clonotypes. For \(i=1,\ldots,N\), let \(\mathbf X_i=(X_{i1},\ldots,X_{ip})\) denote the feature-count vector for sample \(i\), where \(X_{ij}\) is the observed count of feature \(j\), \(j=1,\ldots,p\). The total count \(L_i=\sum_{j=1}^p X_{ij}\) is the library size, or sequencing depth, of sample \(i\). Let \(\mathcal M=\{m_1<\cdots<m_M\}\) denote a prespecified candidate grid of rarefying depths. For each sample \(i\) and each \(m_\ell\in\mathcal M\) satisfying \(m_\ell\le L_i\), the observed count profile is rarefied without replacement to depth \(m_\ell\), and the selected alpha diversity measure is computed. When repeated rarefying is used to reduce variation from the random subsampling step, let \(A_i^{(r)}(m_\ell)\), \(r=1,\ldots,R\), denote the alpha diversity estimate from the \(r^{\text{th}}\) rarefying replicate. We use the replicate average as the alpha diversity estimate at depth \(m_\ell\),
\[
Y_{i\ell}
=
\frac{1}{R}\sum_{r=1}^R A_i^{(r)}(m_\ell).
\]
The resulting sample-by-depth matrix \(\mathbf Y=(Y_{i\ell})\) summarizes how the alpha diversity estimate changes across candidate rarefying depths, with entries defined only when \(m_\ell\le L_i\).

The multi-bin procedure uses the same depth grid \(\mathcal M\) to define both library size thresholds and bin-specific rarefying depths. For a fixed number \(K\) of interior cutpoints, let \(1=\tau_0<\tau_1<\cdots<\tau_K<M\) denote selected grid indices, and let \(\boldsymbol\tau=(\tau_1,\ldots,\tau_K)\) denote the candidate cutpoint configuration. These cutpoints define \(B=K+1\) library size bins, \([m_1,m_{\tau_1}),\ [m_{\tau_1},m_{\tau_2}),\ldots,\ [m_{\tau_K},\infty)\).
Let \(\mathcal I_b(\boldsymbol\tau)\) denote the set of samples assigned to bin \(b\). Within each bin \(b=1,\ldots,B\), the bin-specific rarefying depth is the lower boundary \(m_{\tau_{b-1}}\), and the bin-anchored alpha diversity value for sample \(i\in\mathcal I_b(\boldsymbol\tau)\) is therefore \(\widetilde A_i(\boldsymbol\tau)=Y_{i,\tau_{b-1}}\). This construction preserves comparability within bins while allowing higher-depth samples to be analyzed at larger rarefying depths than would be possible under a single common global rarefying depth.

\subsection{Cutpoint Selection Objective}

The original multi-bin rarefying framework selected library size cutpoints using practical guidance rather than an explicit optimization criterion \citep{li2024multi}. 
We formalize the two main design goals, including reducing residual library size dependence and maintaining sufficiently large bins, through an objective that combines a residual-dependence criterion with a fragmentation penalty.

For a candidate partition \(\boldsymbol\tau\), the first component of the objective measures how much bin-anchored alpha diversity remains associated with the original library size. Within bin \(b\), let \(\ell_i=\log_{10}(L_i)\), where the log transformation reflects that library sizes often vary over orders of magnitude and that residual depth effects are more naturally assessed on a multiplicative scale. Additional covariates, denoted by \(\mathbf Z_i\), may be included to adjust the assessment of residual library size dependence. To measure the additional variation in \(\widetilde A_i(\boldsymbol\tau)\) explained by library size within bin \(b\), after accounting for \(\mathbf Z_i\), we compare two least-squares fits over samples in \(\mathcal I_b(\boldsymbol\tau)\). The reduced model excludes library size,
\[
\widetilde A_i(\boldsymbol\tau)
=
\beta_{0b}
+
\mathbf Z_i^\top\boldsymbol\gamma_b
+
\varepsilon_{ib},
\]
whereas the full model includes the log-library size term,
\[
\widetilde A_i(\boldsymbol\tau)
=
\beta_{0b}
+
\beta_{Lb}\ell_i
+
\mathbf Z_i^\top\boldsymbol\gamma_b
+
\varepsilon_{ib}.
\]
Let \(\mathrm{SSE}_{0b}\) and \(\mathrm{SSE}_{1b}\) denote the residual sums of squares from the reduced and full models, respectively. We quantify residual library size dependence in bin \(b\) by the partial coefficient of determination
\[
D_b
=
1-\frac{\mathrm{SSE}_{1b}}{\mathrm{SSE}_{0b}}.
\]
When no adjustment covariates are included, \(D_b\) reduces to the squared within-bin correlation between \(\widetilde A_i(\boldsymbol\tau)\) and \(\log_{10}(L_i)\). The residual library size dependence of the candidate partition \(\boldsymbol\tau\) is measured by the sample-size-weighted average
\[
Q(\boldsymbol\tau)
=
\sum_{b=1}^{B}
\frac{n_b}{N}D_b,
\]
where \(n_b\) is the number of samples in bin \(b\). The weights \(n_b/N\) give larger bins greater influence in the criterion, matching their larger contribution to the downstream association analysis. Smaller values of \(Q(\boldsymbol\tau)\) indicate weaker residual dependence on the original library size after bin anchoring.

Minimizing \(Q(\boldsymbol\tau)\) alone may favor overly fine partitions, because small bins can reduce within-bin library size variation while also making downstream estimation less stable and inference less powerful. To quantify sample fragmentation, we use a penalty based on the reciprocal bin sizes. 
Let \(n_F=\eta_F N\), where \(0<\eta_F<1\) is a user-specified scaling parameter for the fragmentation penalty. We use \(\eta_F=0.05\) as the default value.
For a candidate partition \(\boldsymbol\tau\) with bin sizes \(n_1,\ldots,n_B\), define
\[
F_{\mathrm{raw}}(\boldsymbol\tau)
=
\frac{1}{B}\sum_{b=1}^{B}\frac{n_F}{n_b}.
\]
This quantity increases as bins become smaller or more unbalanced and, even under balanced partitions, increases with the number of bins. For normalization, we use the hypothetical one-bin partition as the zero-fragmentation baseline with the raw penalty as \(F_0=n_F/N=\eta_F\).
The normalized fragmentation penalty is
\[
F(\boldsymbol\tau)
=
\min\left\{
1,
\max\left[
0,
\frac{F_{\mathrm{raw}}(\boldsymbol\tau)-F_0}{1-F_0}
\right]
\right\}.
\]
With this normalization, the hypothetical one-bin partition has penalty zero, while partitions with additional or highly unbalanced bins receive larger penalties. Since \(Q(\boldsymbol\tau)\) is a weighted average of partial \(R^2\) values and therefore lies in \([0,1]\), \(F(\boldsymbol\tau)\) is truncated to the same range so that the two components enter the objective on comparable scales.

The cutpoint-selection objective is then
\begin{equation}\label{eq:obj}
    J(\boldsymbol\tau)=Q(\boldsymbol\tau)+F(\boldsymbol\tau).    
\end{equation}
Candidate partitions are also required to satisfy the hard minimum-bin-size constraint \(n_b\ge n_{\min}\), for \(b=1,\ldots,B\). Partitions that violate this constraint, or for which the within-bin regression models are not estimable, are treated as infeasible. In the designed objective function, \(Q(\boldsymbol\tau)\) targets the residual library size artifact that motivates multi-bin rarefying, while \(F(\boldsymbol\tau)\) discourages reducing this artifact through excessive sample fragmentation. The hard minimum-bin-size constraint further excludes partitions that would be unsuitable for stable within-bin association analysis.

\subsection{Fixed-\texorpdfstring{\(K\)}{K} and Varying-\texorpdfstring{\(K\)}{K} Cutpoint Selection}

The primary procedure uses a fixed-\(K\) formulation, in which the number of interior cutpoints is prespecified and only their locations are optimized. For a fixed \(K\), the search is over the ordered grid-index vectors \(\boldsymbol\tau\). A cutpoint configuration is feasible if its grid indices are strictly ordered and each induced bin contains at least \(n_{\min}\) samples. We denote the feasible set by
\[
\mathcal T_K =
\left\{
\boldsymbol\tau:
1<\tau_1<\cdots<\tau_K<M,\quad
n_b(\boldsymbol\tau)\ge n_{\min},\ b=1,\ldots,K+1
\right\},
\]
where \(n_b(\boldsymbol\tau)\) is the number of samples in bin \(b\) induced by \(\boldsymbol\tau\). The fixed-\(K\) estimator is
\[
\widehat{\boldsymbol\tau}_K
=
\arg\min_{\boldsymbol\tau\in\mathcal T_K}
J(\boldsymbol\tau).
\]
We use fixed-\(K\) selection as the primary mode because it separates the user-specified partition complexity from the data-adaptive placement of cutpoints. Thus, the user chooses the number of cutpoints, while their locations are determined by the optimization criterion.

As an optional extension, the framework also allows the number of cutpoints to vary. For a prespecified maximum \(K_{\max}\), the varying-\(K\) formulation solves
\[
(\widehat K,\widehat{\boldsymbol\tau})
=
\arg\min_{1\le K\le K_{\max},\ \boldsymbol\tau\in\mathcal T_K}
J(\boldsymbol\tau).
\]
Because the fragmentation penalty discourages additional or highly unbalanced bins, the same objective can be used to compare feasible partitions with different numbers of cutpoints. In the current implementation, the varying-\(K\) search is restricted to \(K\ge 1\), and the hypothetical one-bin partition is used only as the normalization baseline for the fragmentation penalty.

The cutpoint selection problem is a constrained discrete optimization problem over the candidate rarefying-depth grid. For a candidate configuration \(\boldsymbol\tau\), changing one cutpoint can alter the sample partition, the bin-specific rarefying depths, and consequently the value of the objective \(J(\boldsymbol\tau)\) in Equation~\eqref{eq:obj}. These features motivate a derivative-free search over the feasible set \(\mathcal T_K\) of ordered grid-index configurations. We implement this search using the ordered-knot genetic algorithm framework in \texttt{GAReg} \citep{li2026GAReg}, with \texttt{changepointGA} as the computational backend \citep{li2026changepointga}. In the present application, the ordered locations optimized by \texttt{GAReg} are reinterpreted as library size cutpoints on the candidate depth grid. A candidate solution encodes an ordered grid-index vector \(\boldsymbol\tau=(\tau_1,\ldots,\tau_K)\), where \(K\) may be fixed in advance or searched over in the varying-\(K\) formulation, and the selected indices correspond to the cutpoints \(m_{\tau_1},\ldots,m_{\tau_K}\). 

For each candidate solution, the induced bins are constructed, the bin-anchored alpha diversity values are extracted from the rarefying profile matrix \(\mathbf Y\), and the objective \(J(\boldsymbol\tau)\) is evaluated. The ordering machinery in \texttt{GAReg}/\texttt{changepointGA} is used to generate ordered candidate cutpoints, while the minimum-bin-size requirement is enforced through the feasibility check. A candidate configuration is treated as feasible only if every induced bin, including the two boundary bins, contains at least \(n_{\min}\) samples. Configurations that violate this requirement, or for which the within-bin regression models used in \(Q(\boldsymbol\tau)\) are not estimable, are assigned an infeasible objective value and excluded from selection. Conditional on the rarefying profiles \(\mathbf Y\), the objective \(J(\boldsymbol\tau)\) is deterministic. The stochasticity enters only through the genetic algorithm search. We therefore evaluate the stability of the selected cutpoints across random seeds in the simulation study.

\subsection{Association Analysis and Diagnostics}

As shown in Figure~\ref{fig:mbr_workflow}, data-adaptive binning and multi-bin inference are treated as separate stages. The proposed optimization procedure determines the library size cutpoints, whereas the subsequent bin-wise association analysis follows the multi-bin rarefying framework of \citep{li2024multi}. Thus, the cutpoint-selection objective targets residual library size dependence and does not use the biological or ecological exposure of interest.

After the cutpoints are selected, association analysis is performed within each bin using
\[
\widetilde A_i
=
\theta_{0b}
+
\theta_{b}x_i
+
\mathbf W_i^\top\boldsymbol\delta_b
+
e_{ib},
\qquad i\in\mathcal I_b,
\]
where \(x_i\) is the phenotype, exposure, or ecological covariate of interest, and \(\mathbf W_i\) denotes optional covariates for the downstream association model. The bin-specific estimate \(\widehat\theta_b\) targets the association between \(x_i\) and bin-anchored alpha diversity within bin \(b\).

The resulting bin-specific estimates are combined across bins by meta-analysis, as in the original multi-bin rarefying approach \citep{li2024multi}. In the numerical studies, inverse variance weighting (IVW) is used as the primary combination rule:
\[
\widehat\theta_{\mathrm{IVW}}
=
\frac{\sum_{b=1}^{B}\widehat V_b^{-1}\widehat\theta_b}
{\sum_{b=1}^{B}\widehat V_b^{-1}},
\qquad
\widehat{\mathrm{Var}}(\widehat\theta_{\mathrm{IVW}})
=
\left(\sum_{b=1}^{B}\widehat V_b^{-1}\right)^{-1},
\]
where \(\widehat V_b\) is the estimated variance of \(\widehat\theta_b\).

Residual library size dependence is assessed separately after cutpoint selection as a diagnostic analysis. Using the selected bins and same bin-anchored alpha diversity values, the diagnostic model replaces the scientific exposure \(x_i\) with \(\log_{10}(L_i)\). A non-significant residual library size test is interpreted as evidence that no detectable depth association remains under the diagnostic model, rather than as proof of complete independence. For reproducibility, the selected cutpoints, bin sizes, candidate depth grid, and random seed should be reported. The full workflow in Figure~\ref{fig:mbr_workflow} is implemented in the \texttt{MBRarefy} R package. The implementation supports repeated rarefying, fixed-\(K\) and varying-\(K\) cutpoint selection, bin-wise association testing, cross-bin meta-analysis, and residual library size diagnostics.

\section{Simulation Studies}\label{sec:sim}

In this section, we conducted simulation studies to evaluate the proposed framework under heterogeneous library sizes. Rather than considering only a single null and alternative, the design varied three factors, including biological effect strength, group and library size association, and library size dependent underlying diversity. The biological effect controls the true difference in underlying alpha diversity between groups. Group and library size association controls whether the two groups have different library-size distributions. Library size dependent underlying diversity controls whether the set of sequence types available for sampling changes systematically with library size. Varying these factors separately allows us to distinguish true biological association from depth-related artifacts and to evaluate both inferential validity and efficiency.

\subsection{Simulation Design and Competing Methods}

We used an empirical TCR sequence pool construction adapted from Simulation B of \citet{li2024multi}. The original data construction uses two observed TCR repertoires from the cytomegalovirus (CMV) immunosequencing study of \citet{emerson2017immunosequencing} as empirical sequence-frequency sources. These repertoires exhibit substantial dissimilarity in terms of beta diversity. We retained this empirical pool framework while separately controlling for biological perturbation, group and library size association, and library size dependent underlying diversity.

Sequences with counts of 1000 or more were first removed from each source repertoire. We then randomly selected 1,000 sequence types, together with their empirical frequencies, from each source to form two reference pools.
Let \(A\) denote the reference pool derived from the first TCR repertoire and \(A^\prime\) the reference pool derived from the second. Because the two source repertoires were substantially different, \(A\) and \(A^\prime\) represent distinct sequence-frequency sources. The sampling pool for group A consisted of the whole \(A\), whereas that for group B was constructed as \(B(p)=A\cup A^\prime_p\), where \(A^\prime_p\) contains a proportion \(p\) of the 1,000 sequence types in \(A^\prime\), together with their empirical frequencies. 
The parameter \(p\) therefore controls the strength of the biological perturbation between the two groups. When \(p=0\), the two groups share the same sampling pool, and there is no biological group effect. As \(p\) increases, progressively more sequence types from the second reference pool are introduced into the group B pool, producing an increasingly distinct underlying repertoire. We considered \(p\in\{0,0.001,0.002,0.003,0.004,0.005,0.006,0.009,0.05,0.09,0.20,0.30\}\), with \(p=0\) providing the null setting and \(p>0\) representing biological perturbations of increasing strength. For example, \(p=0.003\) adds three sequence types from \(A^\prime\) to the group B sampling pool, while group A continues to be generated from the original pool \(A\).

% make group and library size confounding
For each simulated sample, the original library size \(L_i\) was generated over the range \(10^3\) to \(10^5\) reads. 
We used \(G\in\{0,1\}\) to indicate the absence or presence of group and library size association. When \(G=0\), \(\log_{10}(L_i)\sim \operatorname{Uniform}(3,5)\) independently in the two groups. When \(G=1\), we first generated \(N\) library sizes from the same distribution and sorted them in increasing order. The lower and upper halves of the ordered library sizes were assigned sampling weights \(w_{\mathrm{low}} = (1+\delta_{\mathrm{depth}})/N\) and \(w_{\mathrm{high}} = (1-\delta_{\mathrm{depth}})/N\), respectively. Library sizes were then sampled without replacement according to these weights, with the first \(N/2\) sampled values assigned to group A and the remaining \(N/2\) values assigned to group B. 
We set \(\delta_{\mathrm{depth}}=0.4\), so that lower depth observations received weight \(1.4/N\) and higher-depth observations received weight \(0.6/N\) when group A library sizes were selected. Consequently, group A tended to have smaller library sizes and group B tended to have larger library sizes. More importantly, \(G\) controls the group and library size association independently of the biological group effect controlled by \(p\).

% make library size and alpha diversity association
We used \(D\in\{0,1\}\) to indicate the absence or presence of library size dependent underlying diversity. 
When \(D=1\), sequence types in each underlying repertoire were randomly ordered once, and this ordering was retained across simulated samples. Let \(\gamma(L)\) denote the proportion of the underlying repertoire available for sampling at library size \(L\). For a sample with library size \(L\), approximately the first \(\gamma(L)\) proportion of sequence types in the fixed random ordering were available for sampling. Thus, as \(\gamma(L)\) increased with library size, progressively more sequence types became available. We defined
\[
\gamma(L)=
\begin{cases}
0.80
+
0.05
\dfrac{\log_{10}(L)-\log_{10}(1000)}
      {\log_{10}(3000)-\log_{10}(1000)},
& 1000 \leq L < 3000, \\[10pt]
0.95
+
0.03
\dfrac{\log_{10}(L)-\log_{10}(3000)}
      {\log_{10}(10000)-\log_{10}(3000)},
& 3000 \leq L < 10000, \\[10pt]
0.98
+
0.01
\dfrac{\log_{10}(L)-\log_{10}(10000)}
      {\log_{10}(30000)-\log_{10}(10000)},
& 10000 \leq L < 30000, \\[10pt]
1,
& 30000 \leq L \leq 100000.
\end{cases}
\]
This construction creates designed structural changes in the relationship between library size and underlying accessible diversity at 3,000, 10,000, and 30,000 reads. These locations are therefore used as the reference structural locations in analyses of the adaptive partition. When \(D=0\), we set \(\gamma(L)=1\) for all samples, so that the full potential repertoire was accessible regardless of library size. 

After determining the sequence types available for sample \(i\), their empirical frequencies were retained, and \(L_i\) reads were sampled without replacement to generate the observed count profile. Thus, library size influenced the simulated count profiles in two ways. In all settings, it determined the number of reads sampled. When \(D=1\), it also determined the proportion of the underlying repertoire available for sampling through \(\gamma(L)\).

% simulation set-up

The main simulation followed a factorial design over
\((G,D)\in\{0,1\}^2\), biological perturbation strength \(p\), and total sample size \(N\). The four combinations of \(G\) and \(D\), summarized in Table~\ref{tab:simulation_scenarios}, separately represent the presence or absence of group and library size association and library size-dependent underlying diversity. We considered total sample sizes of \(N=200,400,\) and \(800\), corresponding to 100, 200, and 400 samples per group. For each simulation setup, we generated 300 Monte Carlo replicates. The main simulations used richness as the alpha diversity measure. To avoid additional Monte Carlo variation from repeated subsampling, rarefied richness at each candidate depth was evaluated using its exact hypergeometric expectation.

\begin{table}[!h]
\centering
\caption{Factorial simulation study design.}
\label{tab:simulation_scenarios}
\begin{tabular}{cp{0.71\linewidth}}
\toprule
\((G,D)\) & Interpretation \\
\midrule
\((0,0)\) & Neither group and library size association nor library size dependent diversity \\
\((1,0)\) & Group and library size association only \\
\((0,1)\) & library size dependent diversity only \\
\((1,1)\) & Both group and library size association and library size dependent diversity \\
\bottomrule
\end{tabular}
\end{table}

Five rarefaction strategies were compared. The first was the \emph{no rarefying} baseline, which used alpha diversity computed directly from the original count profiles. The second, \emph{overall rarefying}, rarefied all samples to the common minimum depth of 1,000 reads before testing the group effect. The third comparator was the \emph{prespecified multi-bin} procedure of \citet{li2024multi}, using fixed library size cutpoints at 3,000, 10,000, and 30,000 reads. These cutpoints coincide with the structural changes in the data-generating mechanism when \(D=1\) and therefore provide an oracle structural benchmark for the adaptive procedures. When \(D=0\), they serve only as prespecified benchmark locations rather than true structural cutpoints. Within each bin, samples were rarefied to the lower bin boundary, and the bin-specific group-effect estimates were combined by IVW.

The primary proposed method was \emph{fixed-\(K\) MBRarefy} with \(K=3\), for which the three cutpoint locations were selected adaptively by minimizing the objective in Equation~\eqref{eq:obj}. We also evaluated the optional \emph{varying-\(K\) MBRarefy} extension, which jointly selected the number and locations of cutpoints subject to the minimum-bin-size constraint. For both \emph{MBRarefy} procedures, downstream association analysis followed the same bin-wise estimation and IVW combination used for the prespecified multi-bin comparator.

For the two data adaptive procedures, the default candidate rarefying-depth grid contained 55 depths, spaced every 1,000 reads from 1,000 to 10,000 and every 2,000 reads from 12,000 to 100,000. We set \(n_{\min}=20\) and \(\eta_F=0.05\). The varying-\(K\) procedure searched over \(K\geq1\) up to the maximum feasible value. To reduce optimization error, the fixed-\(K\) genetic algorithm used a population size of 100 and 7,500 generations. The varying-\(K\) island model genetic algorithm used five islands, a population size of 100 per island, 200 migrations, 100 generations per migration, and a convergence limit of 201. For each replicate, we recorded the estimated group effect, its model-based standard error, and the rejection indicator at the 0.05 significance level. For the data adaptive procedures, we additionally recorded \(Q\), \(F\), the total objective \(J\), and the selected cutpoints; for the varying-\(K\) procedure, we also recorded the selected number of cutpoints.

\subsection{Inferential Performance across Biological Effect Strengths}

\begin{figure}[!ht]
  \centering
  \includegraphics[width=\textwidth]{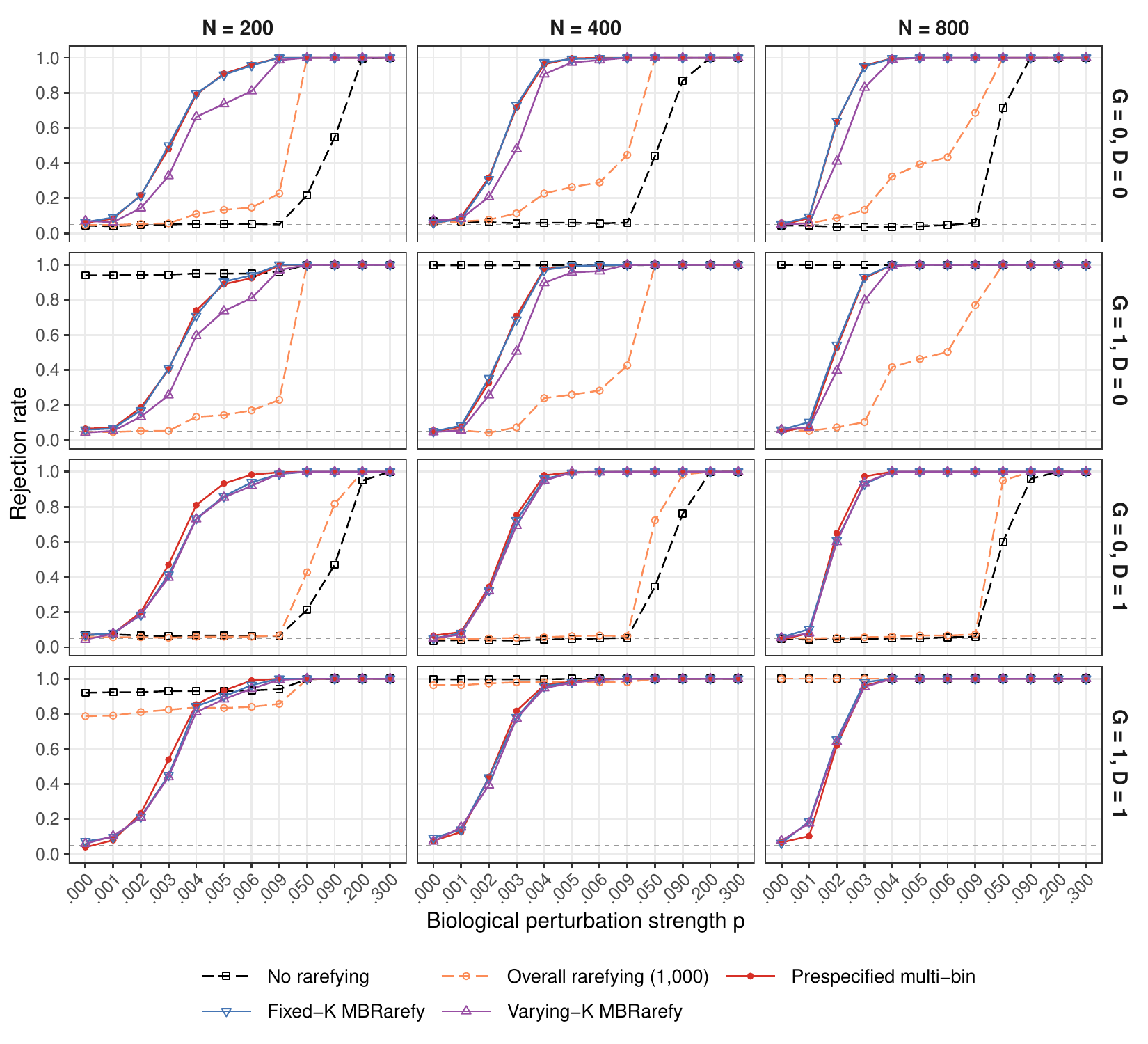}
  \caption{Empirical rejection rates for the five analysis strategies under the factorial simulation design. The horizontal dashed line indicates the nominal 0.05 significance level.}
    \label{fig:inferential_performance}
\end{figure}

We examined the five competing procedures across the factorial simulation settings in Table~\ref{tab:simulation_scenarios}. Figure~\ref{fig:inferential_performance} shows the corresponding empirical rejection rates based on 300 simulation replicates per setting.

We first considered $G=0$, where group membership was independent of library size. 
% overall
Under $p=0$, all five procedures showed rejection rates close to the nominal 0.05 level for both $D=0$ and $D=1$, indicating that library size dependent diversity alone did not induce systematic false positive inflation in the absence of group and library size association.
% separately for D=0 and D=1
As $p$ increased, power increased with both biological perturbation strength and sample size $N$, but clear differences emerged among the competing procedures. Overall rarefying to a single common depth can incur information loss by excluding samples with library sizes below the rarefaction depth and thinning the samples with larger library sizes to the same common depth. The multi-bin procedures use depth specific rarefaction levels, then retaining more samples and preserving more sequencing information from higher-depth samples. This greater information retention translated into substantially higher power across a range of alternatives. 

Among the multi-bin procedures, the prespecified and fixed-$K$ methods generally showed very similar and the highest power under both $D=0$ and $D=1$. In both settings, fixing $K=3$ avoided the additional variability associated with selecting the number of cutpoints. Under $D=1$, the prespecified multi-bin method also benefited from having cutpoints aligned with the designed structural locations, while the adaptive fixed-$K$ procedure estimated their locations from the data and achieved comparable power. In contrast, the varying-$K$ procedure selected both the number and locations of cutpoints, leading to greater partition selection variability and slightly lower power.

We next considered the $G=1$ scenarios, where group membership was associated with library size. The main difference from the $G=0$ settings was in type I error control. When $D=0$, no rarefying produced severe false-positive inflation, whereas overall rarefying to 1,000 reads and the multi-bin procedures maintained rejection rates near the nominal level. When $D=1$, however, overall rarefying was no longer sufficient to control the depth-induced distortion, and both no rarefying and overall rarefying showed substantial inflation under $p=0$. In contrast, the prespecified and adaptive multi-bin procedures largely controlled this inflation, with rejection rates remaining much closer to the nominal level across sample sizes.

Taken together, these results provide no evidence that multi-bin rarefying materially distorted the simulated biological signal. The multi-bin procedures improved control of library size induced distortion while preserving strong power across the alternatives, avoiding the substantial information loss associated with overall rarefying to a single common depth.

\subsection{Adaptive Partition Selection and Structural Recovery}

To more directly assess the performance of data-adaptive bin cutpoint selection, we focused on the $D=1$ scenarios in the factorial simulation design. The data generating process contained the three designed cutpoints at 3000, 10000, and 30000. We evaluated the adaptive procedures under these settings to assess their ability to recover the designed partition.

To quantify joint recovery of the designed partition, we used the changepoint distance implemented in the \texttt{changepointGA} R package \citep{li2026changepointga}.
Let \( \mathcal{C}_0=\{\tau_1,\ldots,\tau_{K_0}\}\) and \(\widehat{\mathcal{C}}=\{\widehat{\tau}_1,\ldots,\widehat{\tau}_{\widehat K}\}\) denote the designed and estimated cutpoint configurations, respectively. The distance is
defined as
\[
d(\mathcal{C}_0,\widehat{\mathcal{C}})
=
|K_0-\widehat K|
+
\min_{\mathbf I}
\sum_{i=1}^{K_0}
\sum_{j=1}^{\widehat K}
\frac{|\tau_i-\widehat{\tau}_j|}{L_{\max}} I_{ij},
\]
where $\mathbf I=(I_{ij})$ defines a one-to-one matching between
$\min(K_0,\widehat K)$ pairs of cutpoints, and $L_{\max}$ is the rarefying depth scale. The first term measures disagreement in the number of cutpoints, whereas the second measures the location discrepancy among optimally matched cutpoints. A distance of zero indicates exact recovery of both the number and locations of the cutpoints. In our simulations, \(\mathcal{C}_0=\{3000,10000,30000\}\), \(K_0=3\), and \(L_{\max}=100000\). Figure~3 reports the average of this distance over the simulation replicates.
The fixed-$K$ procedure maintained very small distances across all biological perturbation strengths, indicating the estimated cutpoint locations remained close to the designed locations when $K$ was fixed at three. 
The varying-$K$ procedure also showed relatively small distances for small and moderate values of $p$, but the distance increased sharply at larger $p$, particularly from $p=0.09$ onward. This pattern was consistent across sample sizes and under both $G=0$ and $G=1$. To disentangle errors in the number and locations of selected cutpoints, Supplementary Figure~S1 reports the proportion of replications selecting three cutpoints. Supplementary Figure~S2 evaluates localization accuracy among replications with $\widehat K=3$ using the mean absolute error on the $\log_{10}$ depth scale,
\(\frac{1}{3}\sum_{j=1}^{3}\left| \log_{10}(\widehat{\tau}_j)-\log_{10}(\tau_j)\right|\).

\begin{figure}[!htbp]
  \centering
  \includegraphics[width=0.9\textwidth]{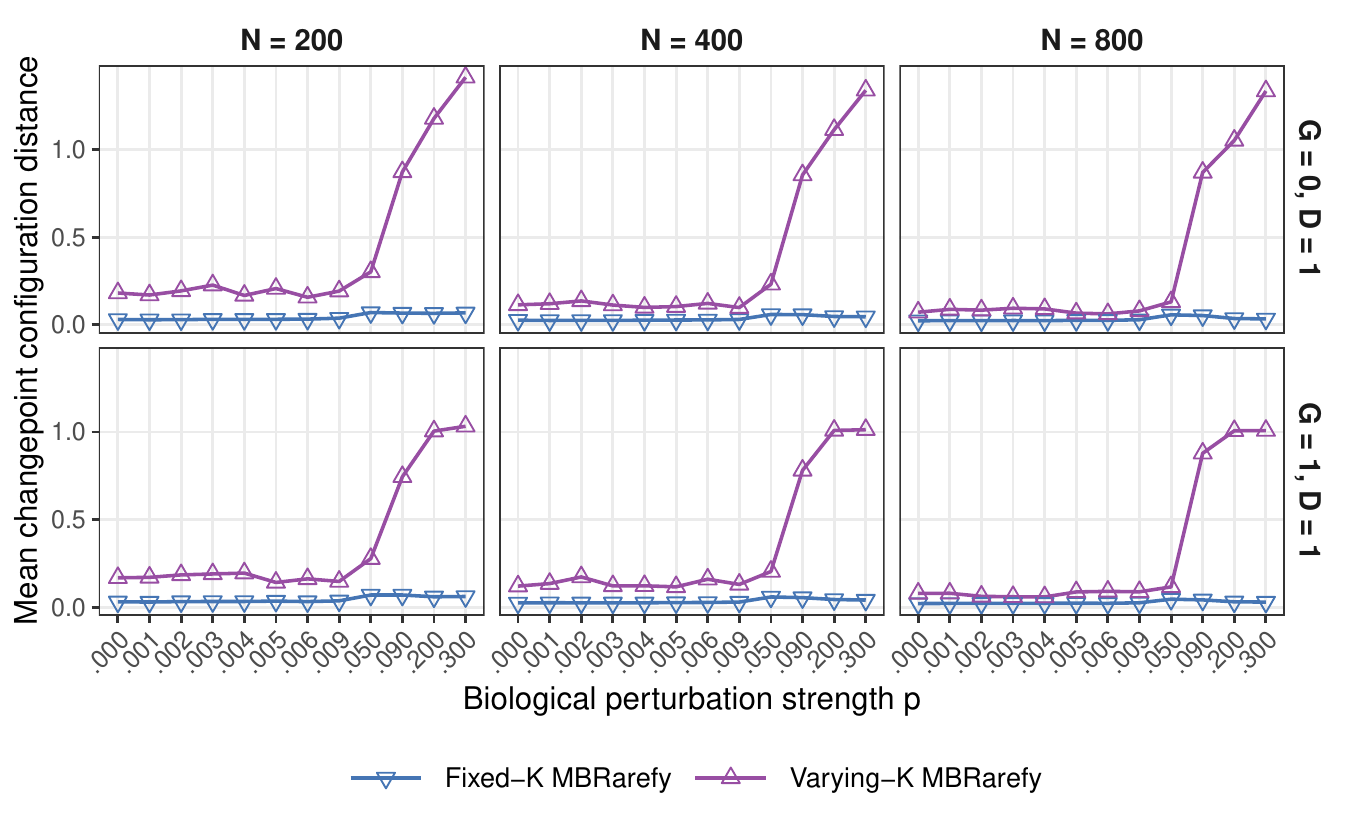}
  \caption{Average distance between the estimated and designed cutpoint configurations under library size-dependent diversity (\(D=1\)). The designed configuration contains three cutpoints at 3,000, 10,000, and 30,000 reads. The distance jointly reflects discrepancies in both the number and locations of selected cutpoints. Smaller values indicate closer agreement with the designed depth structure.}
  \label{fig:structure_recovery_cptdist}
\end{figure}

The decreased performance in cutpoint recovery for the varying-$K$ procedure at large $p$ was associated with the behavior of $Q$, the residual library size dependence component of the partition selection criterion in Equation~\eqref{eq:obj}.
As the biological group difference became stronger, between-group variation contributed increasingly to the total variation in alpha diversity, reducing the proportion of variation attributed to library size. A diagnostic analysis in Section~S1 of the Supplementary Material showed that the group-adjusted criterion remained substantial at large $p$ and recovered the designed three-cutpoint structure, indicating that the underlying depth structure remained present. More importantly, the decline in structural recovery at large $p$ did not correspond to a loss of inferential performance. As shown in Figure~\ref{fig:inferential_performance}, the multi-bin procedures had already achieved high power in these settings. Structural recovery and downstream association performance therefore characterize distinct aspects of the adaptive partitioning procedure. 

Additional analyses in Section~S2 of the Supplementary Material showed that the primary fixed-\(K\) procedure was stable across candidate grids and genetic algorithm initializations, whereas the varying-\(K\) procedure showed greater optimization sensitivity, particularly on the finest candidate grid.

\subsection{Sensitivity to Prespecified-Partition Misspecification}

To emphasize the impact of cutpoint misspecification on downstream inference, we compared the designed prespecified partition with mildly and moderately perturbed configurations. This analysis was restricted to settings with library size dependent diversity ($D=1$). We considered three representative settings at $N=400$ in Table~\ref{tab:partition_misspecification}, including a null setting with group and library size association ($G=1,p=0$), and two alternative settings with $p=0.003$, one without ($G=0$) and one with ($G=1$) group and library size association. The adaptive fixed-$K$ procedure, which estimated three cutpoint locations from the observed depth and diversity relationship, was included as a data-adaptive reference.

\begin{table}[!h]
\centering
\caption{Sensitivity of downstream inference to prespecified partition misspecification under library size dependent diversity ($D=1$). Results are based on 300 Monte Carlo replicates with $N=400$.}
\label{tab:partition_misspecification}
\small
\begin{tabular}{lllrrrr}
\toprule
\multirow{2}{*}{Scenario}
& \multirow{2}{*}{Partition}
& \multirow{2}{*}{Cutpoints}
& Rejection
& Mean
& Mean
& Mean \\
& & 
& rate
& estimate
& SE
& $Q$ \\
\midrule

\multirow{2}{*}{Null}
& Designed
& $(3{,}000,10{,}000,30{,}000)$
& 0.063 & 0.059 & 0.607 & 0.220 \\
& Mild
& $(4{,}000,12{,}000,34{,}000)$
& 0.073 & 0.277 & 0.704 & 0.369 \\
\multirow{2}{*}{($G=1,p=0$)}
& Moderate
& $(5{,}000,16{,}000,40{,}000)$
& 0.110 & 0.517 & 0.739 & 0.467 \\
& Adaptive
& ---
& 0.050 & 0.127 & 0.605 & 0.191 \\
\midrule

\multirow{2}{*}{Alternative}
& Designed
& $(3{,}000,10{,}000,30{,}000)$
& 0.763 & 1.594 & 0.588 & 0.216 \\
& Mild
& $(4{,}000,12{,}000,34{,}000)$
& 0.820 & 1.936 & 0.680 & 0.366 \\
\multirow{2}{*}{($G=0,p=0.003$)}
& Moderate
& $(5{,}000,16{,}000,40{,}000)$
& 0.790 & 2.017 & 0.714 & 0.464 \\
& Adaptive
& ---
& 0.727 & 1.520 & 0.589 & 0.187 \\
\midrule

\multirow{2}{*}{Alternative}
& Designed
& $(3{,}000,10{,}000,30{,}000)$
& 0.760 & 1.642 & 0.614 & 0.217 \\
& Mild
& $(4{,}000,12{,}000,34{,}000)$
& 0.827 & 2.108 & 0.709 & 0.366 \\
\multirow{2}{*}{($G=1,p=0.003$)}
& Moderate
& $(5{,}000,16{,}000,40{,}000)$
& 0.923 & 2.556 & 0.743 & 0.469 \\
& Adaptive
& ---
& 0.720 & 1.609 & 0.613 & 0.189 \\
\bottomrule
\end{tabular}
\end{table}

Across all three settings, misspecification of the cutpoints increased residual library size dependence, with larger departures from the designed partition generally producing larger values of $Q$. Under the null, for example, mean $Q$ increased from 0.220 under the designed partition to 0.467 under moderate misspecification, while the rejection rate increased from 0.063 to 0.110. In contrast, the adaptive fixed-$K$ procedure yielded a lower mean $Q$ of 0.191 and a rejection rate of 0.050.

A similar pattern was observed under the alternatives. Misspecified partitions produced greater residual depth dependence and shifted the estimated biological effect, with the distortion becoming more consequential when group membership was also associated with library size. In the $G=1$ alternative, moderate misspecification increased mean $Q$ from 0.217 to 0.469 and shifted the mean estimated effect from 1.642 to 2.556. The adaptive fixed-$K$ procedure instead produced a mean $Q$ of 0.189 and an estimated effect of 1.609, remaining close to the designed partition.

These results show that poorly chosen prespecified cutpoints can affect more than efficiency. The residual library size dependence can distort both downstream estimation and rejection behavior. Our data-adaptive cutpoint selection reduces reliance on manual partition specification while retaining behavior close to the designed structural benchmark.

\section{Applications}\label{sec:cases}

We applied \texttt{MBRarefy} to a TCR immune repertoire dataset with known CMV serostatus and a wild baboon gut microbiome dataset. Observed richness and Shannon diversity were used as representative alpha diversity outcomes. We compared no rarefying and overall rarefying with the data-adaptive fixed-$K$ and varying-$K$ multi-bin procedures. For the TCR application, we additionally included the prespecified multi-bin procedure of \citet{li2024multi}. Comparisons focused on residual association with original library size, retained sample size, and resulting associations with biologically or ecologically relevant covariates.

\subsection{TCR immune repertoire application}

Following the application in \citet{li2024multi}, we analyzed Cohort~1 of the TCR immunosequencing dataset from \citet{emerson2017immunosequencing}. The original cohort contains TCR-$\beta$ repertoire profiles from 666 subjects with known CMV serostatus, together with demographic information including age and sex. TCR repertoires were obtained by immunosequencing peripheral-blood mononuclear cells, with each sample represented by counts of observed TCR sequences. Restricting the analysis to samples with original library sizes of at least \(10^5\) reads, the minimum candidate rarefaction depth, yielded 663 eligible profiles.

For these 663 eligible profiles, we computed richness and the Shannon index over a grid of \(M=100\) candidate rarefaction depths spanning \(10^5\) to \(10^7\) reads. At each candidate depth not exceeding a sample's original library size, reads were subsampled without replacement 10 times. Richness and the Shannon index were calculated for each rarefied replicate and then averaged across 10 replicates to reduce Monte Carlo variability introduced by rarefaction. The resulting sample-by-depth diversity profiles characterize how the estimated alpha diversity measures vary with library size and provide the input for the \texttt{MBRarefy} data-adaptive partition-selection procedure. 

For comparability, all rarefaction and association analyses were recomputed using the same analysis pipeline. Overall rarefying was performed at \(10^6\) reads. For the prespecified multi-bin analysis, we used the cutpoints \(1, 2, 4, 8,\) and \(10\) million reads from \citet{li2024multi}. For the adaptive procedures, all candidate partitions were required to contain at least 10 samples per bin. The fixed-$K$ procedure was evaluated with \(K=3\) and \(K=5\) interior cutpoints, while the varying-$K$ procedure selected both the number and locations of the cutpoints. For all multi-bin procedures, bin-specific association estimates were combined using IVW.

\begin{table*}[!h]
\centering
\caption{Association results for the TCR immune repertoire application. Entries are estimated regression coefficients with corresponding $P$-values in parentheses. Multi-bin estimates were combined using inverse-variance weighting. All adaptive analyses used a minimum bin size of 10. Bold $P$-values indicate statistical significance at the 0.05 level.}
\label{tab:TCR_application}
\small

\resizebox{\textwidth}{!}{%
\begin{tabular}{lcccccc}
\toprule
& No rarefying
& Overall rarefying
& Prespecified
& Fixed-$K$
& Fixed-$K$
& Varying-$K$ \\
&
& ($10^6$)
& multi-bin (IVW)
& ($K=3$)
& ($K=5$)
& multi-bin \\
\midrule

\multicolumn{7}{c}{\textit{Panel A: Richness}} \\
\midrule

Retained
& \multirow{2}{*}{663}
& \multirow{2}{*}{624}
& \multirow{2}{*}{663}
& \multirow{2}{*}{663}
& \multirow{2}{*}{663}
& \multirow{2}{*}{663} \\
sample size
& & & & & & \\[2pt] \hline

Cutpoints
& --
& --
& \multirow{1}{*}{$1.0, 2.0, 4.0$}
& \multirow{2}{*}{$2.0, 4.3, 7.2$}
& \multirow{2}{*}{$1.9, 3.0, 4.5, 6.3, 8.1$}
& \multirow{2}{*}{$3.2$} \\
($\times 10^6$ reads)
&
&
& \multirow{1}{*}{$8.0, 10.0$}
&
&
& \\[2pt] \hline

\multirow{2}{*}{Age ($\times 10^{-4}$)}
& $-0.151$
& $-0.120$
& $-0.092$
& $-0.043$
& $-0.045$
& $-0.030$ \\
& {\bfseries\boldmath $(7.20\times10^{-7})$}
& {\bfseries\boldmath $(1.52\times10^{-9})$}
& {\bfseries\boldmath $(1.95\times10^{-10})$}
& {\bfseries\boldmath $(2.12\times10^{-8})$}
& {\bfseries\boldmath $(1.46\times10^{-8})$}
& {\bfseries\boldmath $(1.04\times10^{-6})$} \\[2pt] \hline

\multirow{2}{*}{Sex ($\times 10^{-4}$)}
& $2.012$
& $1.538$
& $1.069$
& $0.483$
& $0.607$
& $0.256$ \\
& {\bfseries\boldmath $(1.20\times10^{-2})$}
& {\bfseries\boldmath $(3.65\times10^{-3})$}
& {\bfseries\boldmath $(2.59\times10^{-3})$}
& {\bfseries\boldmath $(2.16\times10^{-2})$}
& {\bfseries\boldmath $(5.17\times10^{-3})$}
& $(1.17\times10^{-1})$ \\[2pt] \hline

\multirow{2}{*}{CMV ($\times 10^{-4}$)}
& $-1.146$
& $-1.950$
& $-1.455$
& $-0.852$
& $-0.760$
& $-0.769$ \\
& $(1.55\times10^{-1})$
& {\bfseries\boldmath $(2.26\times10^{-4})$}
& {\bfseries\boldmath $(9.72\times10^{-5})$}
& {\bfseries\boldmath $(4.71\times10^{-5})$}
& {\bfseries\boldmath $(6.15\times10^{-4})$}
& {\bfseries\boldmath $(1.04\times10^{-6})$} \\[2pt] \hline

\multirow{2}{*}{$\log_{10}$(Library size) ($\times 10^{-4}$)}
& $17.376$
& $7.276$
& $2.946$
& $1.756$
& $1.686$
& $1.959$ \\
& {\bfseries\boldmath $(6.43\times10^{-61})$}
& {\bfseries\boldmath $(1.11\times10^{-15})$}
& {\bfseries\boldmath $(3.77\times10^{-3})$}
& {\bfseries\boldmath $(3.00\times10^{-4})$}
& {\bfseries\boldmath $(1.03\times10^{-3})$}
& {\bfseries\boldmath $(1.71\times10^{-9})$} \\

\midrule

\multicolumn{7}{c}{\textit{Panel B: Shannon index}} \\
\midrule

Retained
& \multirow{2}{*}{663}
& \multirow{2}{*}{624}
& \multirow{2}{*}{663}
& \multirow{2}{*}{663}
& \multirow{2}{*}{663}
& \multirow{2}{*}{663} \\
sample size
& & & & & & \\[2pt] \hline

Cutpoints
& --
& --
& \multirow{1}{*}{$1.0, 2.0, 4.0$}
& \multirow{2}{*}{$2.0, 4.5, 7.1$}
& \multirow{2}{*}{$1.9, 3.2, 4.8, 6.7, 8.4$}
& \multirow{2}{*}{$3.2$} \\
($\times 10^6$ reads)
&
&
& \multirow{1}{*}{$8.0, 10.0$}
&
&
& \\[2pt] \hline

\multirow{2}{*}{Age}
& $-0.0226$
& $-0.0211$
& $-0.0211$
& $-0.0206$
& $-0.0208$
& $-0.0215$ \\
& {\bfseries\boldmath $(2.25\times10^{-14})$}
& {\bfseries\boldmath $(2.64\times10^{-14})$}
& {\bfseries\boldmath $(1.45\times10^{-16})$}
& {\bfseries\boldmath $(7.31\times10^{-16})$}
& {\bfseries\boldmath $(5.14\times10^{-16})$}
& {\bfseries\boldmath $(6.70\times10^{-17})$} \\[2pt] \hline

\multirow{2}{*}{Sex}
& $0.233$
& $0.212$
& $0.201$
& $0.202$
& $0.225$
& $0.203$ \\
& {\bfseries\boldmath $(2.93\times10^{-3})$}
& {\bfseries\boldmath $(4.45\times10^{-3})$}
& {\bfseries\boldmath $(3.97\times10^{-3})$}
& {\bfseries\boldmath $(3.51\times10^{-3})$}
& {\bfseries\boldmath $(1.28\times10^{-3})$}
& {\bfseries\boldmath $(3.84\times10^{-3})$} \\[2pt] \hline

\multirow{2}{*}{CMV}
& $-0.730$
& $-0.758$
& $-0.736$
& $-0.721$
& $-0.734$
& $-0.733$ \\
& {\bfseries\boldmath $(8.56\times10^{-22})$}
& {\bfseries\boldmath $(2.65\times10^{-26})$}
& {\bfseries\boldmath $(1.84\times10^{-30})$}
& {\bfseries\boldmath $(2.92\times10^{-29})$}
& {\bfseries\boldmath $(3.33\times10^{-30})$}
& {\bfseries\boldmath $(1.14\times10^{-29})$} \\[2pt] \hline

\multirow{2}{*}{$\log_{10}$(Library size)}
& $0.762$
& $0.418$
& $-0.196$
& $-0.048$
& $0.153$
& $0.116$ \\
& {\bfseries\boldmath $(1.28\times10^{-11})$}
& {\bfseries\boldmath $(1.40\times10^{-3})$}
& $(6.12\times10^{-1})$
& $(8.68\times10^{-1})$
& $(6.48\times10^{-1})$
& $(5.19\times10^{-1})$ \\

\bottomrule
\end{tabular}
}

\end{table*}

Table~\ref{tab:TCR_application} summarizes the association analysis results. Before rarefaction, both richness and the Shannon index showed strong associations with original library size. Overall rarefying to $10^6$ reads reduced this dependence but retained only 624 of the 663 eligible samples, and residual library size associations remained statistically detectable for both diversity measures. All multi-bin procedures retained the full set of 663 eligible samples. For richness, the estimated coefficient for \(\log_{10}\) library size, reported on the \(10^{-4}\) scale, decreased from 17.376 without rarefying to 2.946 under the prespecified multi-bin procedure, 1.756 under fixed \(K=3\), 1.686 under fixed \(K=5\), and 1.959 under the varying-$K$ procedure. Although residual library size dependence remained statistically detectable for richness, its magnitude was substantially reduced. For the Shannon index, the residual library size association was not statistically significant under any of the multi-bin procedures.

The adaptive procedures produced distinct partition designs for the two diversity measures. With \(K=3\), the selected cutpoints were \(2.0\), \(4.3\), and \(7.2\) million reads for richness and \(2.0\), \(4.5\), and \(7.1\) million reads for the Shannon index. The varying-$K$ procedure selected a substantially simpler design, with a single cutpoint at \(3.2\) million reads for both diversity measures. The components of the partition-selection objective for the adaptive solutions are summarized in Supplementary Table~S3. Although the fixed-$K$ solutions achieved smaller values of the residual-dependence criterion \(Q\), the varying-$K$ solutions incurred substantially smaller fragmentation penalties and therefore smaller overall objective values. For richness, \(J\) decreased from 0.181 for fixed \(K=3\) to 0.109 for varying \(K\). For the Shannon index, it decreased from 0.170 to 0.075. Increasing the fixed number of cutpoints from three to five produced only a modest reduction in \(Q\) for richness and essentially no improvement for Shannon diversity, while substantially increasing the fragmentation penalty and therefore the overall objective value.

The estimated CMV associations further illustrate the impact of library size related distortion. For richness, the CMV association estimate, reported on the \(\times10^{-4}\) scale used in Table~\ref{tab:TCR_application}, was -1.146 without rarefying and was not statistically significant (\(P=0.155\)). All rarefaction analyses yielded statistically significant negative CMV associations. For the Shannon index, the estimated CMV association remained consistently negative and highly significant across all rarefaction procedures. Age was negatively associated with both diversity measures across methods, and the estimated association with sex was generally positive.

Because CMV status was itself modestly associated with original library size, we additionally examined the sensitivity of the estimated CMV association to adjustment for \(\log_{10}\) library size in Supplementary Table~S4. The unrarefied richness estimate was highly sensitive to this additional adjustment, whereas the corresponding estimates were substantially more stable after rarefaction, particularly under the multi-bin procedures. For the Shannon index, the CMV association was generally stable across all rarefaction strategies and showed only minor changes after additional library size adjustment. These results indicate that multi-bin rarefying substantially reduced the sensitivity of the estimated CMV association to library size adjustment, although some residual library size dependence remained for richness.

\subsection{Wild baboon gut microbiome application}

For the microbiome application, we analyzed the wild baboon gut microbiome dataset available through \texttt{microbiomeDataSets} \citep{microbiomeDataSets}. The full dataset contains 16,234 16S rRNA gut microbiome profiles from 585 wild baboons sampled over 14 years. To construct an independent subject-level example, we retained one profile per baboon by selecting the sample with the largest library size, yielding 585 independent samples.

For these 585 profiles, we computed richness and the Shannon index over a grid of 141 candidate rarefaction depths ranging from 10,000 to 150,000 reads in increments of 1,000 reads. At each candidate depth not exceeding a sample's original library size, reads were subsampled without replacement 10 times, and the resulting diversity estimates were averaged across replicates. We evaluated overall rarefying at 50,000, 69,000, 82,000, and 91,000 reads to illustrate the trade-off between residual library size dependence and sample retention. For the adaptive procedures, all candidate partitions were required to contain at least 10 samples per bin. The fixed-$K$ procedure was evaluated with $K=3$ and $K=5$ interior cutpoints, while the varying-$K$ procedure jointly selected the number and locations of the cutpoints. Bin-specific association estimates were combined using IVW.

\begin{table*}[!h]
\centering
\caption{Association results for the wild baboon gut microbiome application. Entries are estimated regression coefficients with corresponding $P$-values in parentheses.  Multi-bin estimates were combined using IVW. Bold $P$-values indicate statistical significance at the 0.05 level.}
\label{tab:baboon_application}
\small

\resizebox{\textwidth}{!}{%
\begin{tabular}{lcccccccc}
\toprule
& No rarefying
& \multicolumn{4}{c}{Overall rarefying}
& Fixed-$K$
& Fixed-$K$
& Varying-$K$ \\
\cmidrule(lr){3-6}
&
& 50k
& 69k
& 82k
& 91k
& ($K=3$)
& ($K=5$)
& multi-bin \\
\midrule

\multicolumn{9}{c}{\textit{Panel A: Richness}} \\
\midrule

Retained
& \multirow{2}{*}{585}
& \multirow{2}{*}{530}
& \multirow{2}{*}{404}
& \multirow{2}{*}{289}
& \multirow{2}{*}{236}
& \multirow{2}{*}{585}
& \multirow{2}{*}{585}
& \multirow{2}{*}{585} \\
sample size
& & & & & & & & \\[2pt] \hline

Cutpoints
& --
& --
& --
& --
& --
& $58,\ 79,\ 100$
& $58,\ 69,\ 81,\ 96,\ 114$
& $87$ \\
($\times 10^3$ reads)
& & & & & & & & \\[2pt] \hline

\multirow{2}{*}{Age}
& $-0.448$
& $-0.963$
& $-1.169$
& $-0.762$
& $-0.756$
& $-1.296$
& $-1.306$
& $-1.207$ \\
& $(3.78\times10^{-1})$
& $(6.02\times10^{-2})$
& $(5.44\times10^{-2})$
& $(2.93\times10^{-1})$
& $(3.34\times10^{-1})$
& {\bfseries\boldmath $(4.28\times10^{-3})$}
& {\bfseries\boldmath $(3.92\times10^{-3})$}
& {\bfseries\boldmath $(7.79\times10^{-3})$} \\[2pt] \hline

\multirow{2}{*}{Sex}
& $-10.084$
& $-13.349$
& $-12.510$
& $-10.087$
& $-6.068$
& $-7.638$
& $-6.465$
& $-9.789$ \\
& $(5.29\times10^{-2})$
& {\bfseries\boldmath $(1.04\times10^{-2})$}
& {\bfseries\boldmath $(4.15\times10^{-2})$}
& $(1.63\times10^{-1})$
& $(4.40\times10^{-1})$
& $(9.74\times10^{-2})$
& $(1.55\times10^{-1})$
& {\bfseries\boldmath $(3.23\times10^{-2})$} \\[2pt] \hline

\multirow{2}{*}{Group size}
& $0.107$
& $0.153$
& $0.261$
& $0.275$
& $0.434$
& $0.221$
& $0.218$
& $0.161$ \\
& $(3.31\times10^{-1})$
& $(1.74\times10^{-1})$
& $(5.51\times10^{-2})$
& $(8.39\times10^{-2})$
& {\bfseries\boldmath $(1.48\times10^{-2})$}
& {\bfseries\boldmath $(2.32\times10^{-2})$}
& {\bfseries\boldmath $(2.38\times10^{-2})$}
& $(9.56\times10^{-2})$ \\[2pt] \hline

\multirow{2}{*}{$\log_{10}$(Library size)}
& $140.290$
& $108.888$
& $100.396$
& $108.033$
& $81.821$
& $74.450$
& $87.251$
& $125.993$ \\
& {\bfseries\boldmath $(3.72\times10^{-22})$}
& {\bfseries\boldmath $(1.34\times10^{-8})$}
& {\bfseries\boldmath $(2.69\times10^{-4})$}
& {\bfseries\boldmath $(3.63\times10^{-3})$}
& $(5.60\times10^{-2})$
& {\bfseries\boldmath $(6.00\times10^{-3})$}
& {\bfseries\boldmath $(3.41\times10^{-3})$}
& {\bfseries\boldmath $(1.57\times10^{-11})$} \\

\midrule

\multicolumn{9}{c}{\textit{Panel B: Shannon index}} \\
\midrule

Retained
& \multirow{2}{*}{585}
& \multirow{2}{*}{530}
& \multirow{2}{*}{404}
& \multirow{2}{*}{289}
& \multirow{2}{*}{236}
& \multirow{2}{*}{585}
& \multirow{2}{*}{585}
& \multirow{2}{*}{585} \\
sample size
& & & & & & & & \\[2pt] \hline

Cutpoints
& --
& --
& --
& --
& --
& $62,\ 82,\ 107$
& $58,\ 71,\ 84,\ 101,\ 121$
& $84$ \\
($\times 10^3$ reads)
& & & & & & & & \\[2pt] \hline

\multirow{2}{*}{Age}
& $-0.0078$
& $-0.0114$
& $-0.0103$
& $-0.0066$
& $-0.0045$
& $-0.0094$
& $-0.0076$
& $-0.0073$ \\
& $(9.17\times10^{-2})$
& {\bfseries\boldmath $(1.87\times10^{-2})$}
& $(7.18\times10^{-2})$
& $(3.39\times10^{-1})$
& $(5.55\times10^{-1})$
& {\bfseries\boldmath $(4.25\times10^{-2})$}
& $(1.04\times10^{-1})$
& $(1.19\times10^{-1})$ \\[2pt] \hline

\multirow{2}{*}{Sex}
& $-0.115$
& $-0.126$
& $-0.146$
& $-0.127$
& $-0.080$
& $-0.119$
& $-0.117$
& $-0.117$ \\
& {\bfseries\boldmath $(1.55\times10^{-2})$}
& {\bfseries\boldmath $(1.07\times10^{-2})$}
& {\bfseries\boldmath $(1.09\times10^{-2})$}
& $(6.52\times10^{-2})$
& $(2.98\times10^{-1})$
& {\bfseries\boldmath $(1.06\times10^{-2})$}
& {\bfseries\boldmath $(1.16\times10^{-2})$}
& {\bfseries\boldmath $(1.33\times10^{-2})$} \\[2pt] \hline

\multirow{2}{*}{Group size}
& $0.0015$
& $0.0015$
& $0.0021$
& $0.0019$
& $0.0035$
& $0.0013$
& $0.0013$
& $0.0014$ \\
& $(1.31\times10^{-1})$
& $(1.63\times10^{-1})$
& $(9.84\times10^{-2})$
& $(2.11\times10^{-1})$
& {\bfseries\boldmath $(4.78\times10^{-2})$}
& $(1.92\times10^{-1})$
& $(1.90\times10^{-1})$
& $(1.59\times10^{-1})$ \\[2pt] \hline

\multirow{2}{*}{$\log_{10}$(Library size)}
& $0.058$
& $-0.399$
& $-0.691$
& $-0.638$
& $-0.724$
& $0.231$
& $0.068$
& $0.349$ \\
& $(6.72\times10^{-1})$
& {\bfseries\boldmath $(3.04\times10^{-2})$}
& {\bfseries\boldmath $(7.80\times10^{-3})$}
& $(7.26\times10^{-2})$
& $(8.45\times10^{-2})$
& $(4.59\times10^{-1})$
& $(8.54\times10^{-1})$
& $(8.93\times10^{-2})$ \\

\bottomrule
\end{tabular}
}

\end{table*}

Table~\ref{tab:baboon_application} summarizes the resulting association analyses. For richness, the unrarefied data showed a strong positive association with original library size, with an estimated coefficient of 140.290 for $\log_{10}$ library size ($P=3.72\times10^{-22}$). Overall rarefying reduced the magnitude of this association, but the residual library size effect remained statistically significant at 50k, 69k, and 82k reads. At 91k reads, the association was no longer statistically significant ($P=0.056$), but only 236 of the 585 samples were retained. In contrast, all adaptive multi-bin procedures retained the full set of 585 samples. The corresponding library size coefficients were 74.450 for fixed $K=3$, 87.251 for fixed $K=5$, and 125.993 for the varying-$K$ procedure. Thus, the fixed-$K$ procedures substantially reduced the magnitude of residual library size dependence for richness, although statistically detectable dependence remained.

The pattern was different for the Shannon index. The unrarefied Shannon index was not significantly associated with original library size, whereas overall rarefying at 50k and 69k reads produced significant negative residual associations. At 82k and 91k reads, these associations were no longer statistically significant, but at the cost of retaining only 289 and 236 samples, respectively. All three adaptive multi-bin procedures retained all 585 samples and yielded non-significant residual library size associations for the Shannon index. These results illustrate that the effect of rarefying on residual depth dependence can differ across diversity measures and reinforce the importance of reporting a post-rarefaction library size diagnostic rather than assuming that rarefying removes depth dependence automatically.

The adaptive procedures also produced distinct partition designs across the two diversity measures. For richness, the fixed-$K=3$ procedure selected cutpoints at 58k, 79k, and 100k reads, whereas fixed $K=5$ selected 58k, 69k, 81k, 96k, and 114k reads. The varying-$K$ procedure selected a substantially simpler partition with a single cutpoint at 87k reads. For the Shannon index, the corresponding fixed-$K=3$ cutpoints were 62k, 82k, and 107k reads, fixed $K=5$ selected 58k, 71k, 84k, 101k, and 121k reads, and varying-$K$ selected a single cutpoint at 84k reads.

The objective components in Supplementary Table~S5 illustrate the expected trade-off between residual depth dependence and fragmentation. Increasing fixed \(K\) from three to five reduced \(Q\), particularly for the Shannon index, but the additional fragmentation caused the overall objective \(J\) to increase by 52.0\% for richness and 57.5\% for Shannon diversity. In contrast, the varying-\(K\) procedure favored simpler partitions with larger \(Q\) but substantially smaller fragmentation penalties, reducing \(J\) by 31.8\% and 55.0\%, respectively, relative to fixed \(K=3\). These lower objective values reflect the intended balance between residual depth dependence and partition complexity rather than uniformly smaller downstream library-size associations.

Associations with the ecological covariates varied across diversity measures and rarefying designs. Under the adaptive procedures, age was generally negatively associated with diversity, whereas the sex and group-size associations were less consistent across methods. In contrast, the overall-rarefying analyses showed greater variation in statistical significance as the rarefying depth increased and the retained sample size decreased.

Together, the two applications show that \texttt{MBRarefy} can reduce the magnitude of residual library size dependence while retaining substantially more samples and preserving interpretable biological or ecological associations. At the same time, the results show that the extent of residual dependence is diversity-measure and partition specific, underscoring the importance of reporting the post-\texttt{MBRarefy} library size diagnostic alongside the selected partition.

\section{Discussion}\label{sec:Discussion}

\texttt{MBRarefy} extends the multi-bin rarefying framework of \cite{li2024multi} by replacing ad hoc library-size cutpoints with a reproducible data-adaptive design criterion. The objective combines residual within-bin depth dependence, \(Q\), with a fragmentation penalty, \(F\), while a hard minimum-bin-size constraint excludes unstable partitions. This formulation explicitly balances reduction of residual depth dependence against unnecessary sample fragmentation.

The simulation results clarify this statistical trade-off. Under confounded null settings, no rarefying and global minimum-depth rarefying produced substantial false positive inflation, whereas the fixed-\(K\) multi-bin procedure maintained rejection rates near the nominal level. Under alternatives without group and library-size confounding, fixed-\(K\) retained substantially greater power than global rarefying and performed similarly to the prespecified multi-bin benchmark. Sensitivity analyses further showed that fixed-\(K\) was stable across candidate grids and random initializations, whereas varying-\(K\) showed greater sensitivity to optimization on very dense grids. These results support fixed-\(K\) as the primary procedure, with varying-\(K\) serving as an optional extension when partition complexity itself is of interest.

The two applications illustrate the same balance between residual depth dependence and information retention. Adaptive multi-bin rarefying retained all eligible samples in both datasets, whereas higher single rarefying depths excluded substantial numbers of samples. Residual library-size dependence was generally reduced, although not uniformly eliminated, particularly for richness. The application results also emphasize that the partition objective is distinct from the final post-rarefaction library-size diagnostic. A lower \(J\) reflects a more favorable trade-off between residual dependence and fragmentation, not necessarily a smaller downstream library-size coefficient or larger \(P\)-value.

Several limitations and extensions remain. Rarefying standardizes sampling effort but does not correct feature-specific amplification or sequencing biases, so residual library-size diagnostics should still be reported after normalization. In addition, the selected cutpoints can depend on the alpha diversity measure, as observed in the applications. This motivates development of shared or multivariate partition criteria that select common library-size thresholds across multiple diversity measures. Finally, the varying-\(K\) procedure should be viewed as an optional computational extension rather than a universally reliable model-selection rule on arbitrarily dense candidate grids. Overall, \texttt{MBRarefy} provides a reproducible approach for selecting multi-bin rarefying designs that can reduce depth-induced distortion while retaining substantially more sequencing information than conventional single-depth rarefying.

%%%%%%%%%%%%%%%%%%%%%%%%%%%%%%%%%%%%%%%%%%%%%%%%%%%%%%%%
%%%%%%%%%%%%%%%%%%%%%%%%%%%%%%%%%%%%%%%%%%%%%%%%%%%%%%%%%

\section*{Software availability}

The \texttt{MBRarefy} R package, documentation, and illustrative analysis examples are available at \url{https://github.com/mli171/MBRarefy}.

\section*{Data availability statement}

The TCR immunosequencing data analyzed in this study are publicly available from the original study of \citet{emerson2017immunosequencing}. The wild baboon gut microbiome data are publicly available through the \texttt{microbiomeDataSets} R package. 

\section*{Disclosure statement}
No potential conflict of interest was reported by the author.

\section*{Funding}
This research was supported by the Louisiana Board of Regents Support Fund (BoRSF) Research Competitiveness Subprogram, LEQSF(2025-28)-RD-A-18.

\section*{Supplementary material}
Additional simulation results, sensitivity analyses, optimization-stability results, and application-specific partition summaries are provided in the Supplementary Material.

\bibliographystyle{plainnat}
\bibliography{references}

@article{aird2011analyzing,
  author  = {Aird, Daniel and Ross, Michael G. and Chen, Wei-Sheng and
             Danielsson, Maxwell and Fennell, Timothy and Russ, Carsten and
             Jaffe, David B. and Nusbaum, Chad and Gnirke, Andreas},
  title   = {Analyzing and minimizing {PCR} amplification bias in
             {Illumina} sequencing libraries},
  journal = {Genome Biology},
  volume  = {12},
  number  = {2},
  pages   = {R18},
  year    = {2011},
  doi     = {10.1186/gb-2011-12-2-r18}
}

@article{benjamini2012summarizing,
  title={Summarizing and correcting the {GC} content bias in high-throughput sequencing},
  author={Benjamini, Yuval and Speed, Terence P},
  journal={Nucleic acids research},
  volume={40},
  number={10},
  pages={e72--e72},
  year={2012},
  publisher={Oxford University Press}
}

@article{cameron2021enhancing,
  title={Enhancing diversity analysis by repeatedly rarefying next generation sequencing data describing microbial communities},
  author={Cameron, Ellen S and Schmidt, Philip J and Tremblay, Benjamin J-M and Emelko, Monica B and M{\"u}ller, Kirsten M},
  journal={Scientific reports},
  volume={11},
  number={1},
  pages={22302},
  year={2021},
  publisher={Nature Publishing Group UK London}
}

@article{emerson2017immunosequencing,
  title={{Immunosequencing identifies signatures of cytomegalovirus exposure history and HLA-mediated effects on the T cell repertoire}},
  author={Emerson, Ryan O and DeWitt, William S and Vignali, Marissa and Gravley, Jenna and Hu, Joyce K and Osborne, Edward J and Desmarais, Cindy and Klinger, Mark and Carlson, Christopher S and Hansen, John A and others},
  journal={Nature genetics},
  volume={49},
  number={5},
  pages={659--665},
  year={2017},
  publisher={Nature Publishing Group}
}

@article{gloor2017microbiome,
  title={Microbiome datasets are compositional: and this is not optional},
  author={Gloor, Gregory B and Macklaim, Jean M and Pawlowsky-Glahn, Vera and Egozcue, Juan J},
  journal={Frontiers in microbiology},
  volume={8},
  pages={2224},
  year={2017},
  publisher={Frontiers Media SA}
}

@article{hurlbert1971nonconcept,
  title={The nonconcept of species diversity: a critique and alternative parameters},
  author={Hurlbert, Stuart H},
  journal={Ecology},
  volume={52},
  number={4},
  pages={577--586},
  year={1971},
  publisher={Wiley Online Library}
}

@article{hu2022rarefaction,
  title={A rarefaction-without-resampling extension of PERMANOVA for testing presence--absence associations in the microbiome},
  author={Hu, Yi-Juan and Satten, Glen A},
  journal={Bioinformatics},
  volume={38},
  number={15},
  pages={3689--3697},
  year={2022},
  publisher={Oxford University Press}
}

@article{kers2022power,
  title={The power of microbiome studies: some considerations on which alpha and beta metrics to use and how to report results},
  author={Kers, Jannigje Gerdien and Saccenti, Edoardo},
  journal={Frontiers in microbiology},
  volume={12},
  pages={796025},
  year={2022},
  publisher={Frontiers Media SA}
}

@article{lang2018t,
  title={{T cell receptor repertoire among women who cleared and failed to clear cervical human papillomavirus infection: An exploratory proof-of-principle study}},
  author={Lang Kuhs, Krystle A and Lin, Shih-Wen and Hua, Xing and Schiffman, Mark and Burk, Robert D and Rodriguez, Ana Cecilia and Herrero, Rolando and Abnet, Christian C and Freedman, Neal D and Pinto, Ligia A and others},
  journal={PLoS One},
  volume={13},
  number={1},
  pages={e0178167},
  year={2018},
  publisher={Public Library of Science San Francisco, CA USA}
}

@article{li2024multi,
  title={{A multi-bin rarefying method for evaluating alpha diversities in TCR sequencing data}},
  author={Li, Mo and Hua, Xing and Li, Shuai and Wu, Michael C and Zhao, Ni},
  journal={Bioinformatics},
  volume={40},
  number={7},
  pages={btae431},
  year={2024},
  publisher={Oxford University Press}
}

@article{li2026changepointga,
  author  = {Li, Mo and Lu, QiQi},
  title   = {{changepointGA: An R Package for Fast Changepoint Detection via Genetic Algorithms}},
  journal = {The R Journal},
  year    = {2026},
  url     = {https://journal.r-project.org/articles/RJ-2026-058/}
}

@article{li2026GAReg,
  title={{GAR}eg: {G}enetic {A}lgorithms in {R}egression},
  author={Li, Mo and Lu, QiQi and Lund, Robert and Shi, Xueheng},
  journal={arXiv preprint arXiv:2603.14801},
  year={2026}
}

@manual{microbiomeDataSets,
  title = {{microbiomeDataSets: Experiment Hub based microbiome datasets}},
  author = {Leo Lahti and Felix Ernst and Sudarshan Shetty},
  year = {2025},
  note = {R package version 1.19.0},
  url = {https://bioconductor.org/packages/microbiomeDataSets},
  doi = {10.18129/B9.bioc.microbiomeDataSets}
}

@article{mika2025comprehensive,
  title={A comprehensive evaluation of diversity measures for {TCR} repertoire profiling},
  author={Mika, Justyna and Polanska, Alicja and Blenman, Kim RM and Pusztai, Lajos and Polanska, Joanna and Cand{\'e}ias, Serge and Marczyk, Michal},
  journal={BMC biology},
  volume={23},
  number={1},
  pages={133},
  year={2025},
  publisher={Springer}
}

@article{mcknight2019methods,
  title={Methods for normalizing microbiome data: an ecological perspective},
  author={McKnight, Donald T and Huerlimann, Roger and Bower, Deborah S and Schwarzkopf, Lin and Alford, Ross A and Zenger, Kyall R},
  journal={Methods in Ecology and Evolution},
  volume={10},
  number={3},
  pages={389--400},
  year={2019},
  publisher={Wiley Online Library}
}

@article{roswell2021conceptual,
  title={A conceptual guide to measuring species diversity},
  author={Roswell, Michael and Dushoff, Jonathan and Winfree, Rachael},
  journal={Oikos},
  volume={130},
  number={3},
  pages={321--338},
  year={2021},
  publisher={Wiley Online Library}
}

@article{ross2013characterizing,
  author  = {Ross, Michael G. and Russ, Carsten and Costello, Maura and
             Hollinger, Andrew and Lennon, Niall J. and Hegarty, Ryan and
             Nusbaum, Chad and Jaffe, David B.},
  title   = {Characterizing and measuring bias in sequence data},
  journal = {Genome Biology},
  volume  = {14},
  number  = {5},
  pages   = {R51},
  year    = {2013},
  doi     = {10.1186/gb-2013-14-5-r51}
}

@article{weiss2017normalization,
  title={Normalization and microbial differential abundance strategies depend upon data characteristics},
  author={Weiss, Sophie and Xu, Zhenjiang Zech and Peddada, Shyamal and Amir, Amnon and Bittinger, Kyle and Gonzalez, Antonio and Lozupone, Catherine and Zaneveld, Jesse R and V{\'a}zquez-Baeza, Yoshiki and Birmingham, Amanda and others},
  journal={Microbiome},
  volume={5},
  number={1},
  pages={27},
  year={2017},
  publisher={Springer}
}

@article{willis2019rarefaction,
  title={Rarefaction, alpha diversity, and statistics},
  author={Willis, Amy D},
  journal={Frontiers in microbiology},
  volume={10},
  pages={2407},
  year={2019},
  publisher={Frontiers Media SA}
}

\end{document}

%\maketitle
\title{\Large Supplementary Materials of ``Data-Adaptive Multi-Bin Rarefying for Alpha Diversity Association Analysis under Heterogeneous Library Sizes''}
\vspace{0.5cm}
\author{Mo Li\thanks{CONTACT: Mo Li;  Email: mo.li@louisiana.edu}  }
\date{%
Department of Mathematics, University of Louisiana at Lafayette, Lafayette, LA 70504, USA\\
% [2ex]%
% \today
}

\maketitle

\section{Diagnostic Analysis of High-Effect Structural Recovery}

To investigate the deterioration in structural recovery at large biological perturbations $p$, we performed an additional diagnostic analysis under library-size dependent diversity ($D=1$). We considered $G=0$ and $G=1$, with $p\in\{0,0.003,0.30\}$ and $N=400$. For each setting, the data-generating process and genetic-algorithm settings were identical to those used in the main simulation study.

The diagnostic focused on the residual library-size dependence component \(Q(\tau)\) of the partition-selection criterion. In the main simulation study, no adjustment covariates were included when evaluating \(Q(\tau)\). Thus, within each bin, the reduced model contained only an intercept, whereas the full model additionally included \(\log_{10}(L)\). The resulting partial \(R^2\) therefore measured the marginal within-bin
association between bin-anchored alpha diversity and original library-size.

For diagnostic purposes in this subsection only, we recomputed the criterion after including the biological group indicator \(G\) as an adjustment covariate. Specifically, within each bin, the reduced model was
\[
\widetilde A_i(\tau) = \beta_{0b} + \beta_{Gb} G_i + \varepsilon_{ib},
\]
whereas the full model was
\[
\widetilde A_i(\tau) = \beta_{0b} + \beta_{Gb} G_i + \beta_{Lb}\log_{10}(L_i) + \varepsilon_{ib}.
\]
The resulting partial \(R^2\) therefore quantified the additional variation in bin-anchored alpha diversity explained by library-size after accounting for the biological group effect.

\begin{table}[!htbp]
\centering
\caption{Diagnostic comparison of marginal and group-adjusted residual library-size dependence at the designed partition $(3{,}000,10{,}000,30{,}000)$ under library-size dependent diversity ($D=1$). Values are averaged over three simulated datasets per setting.}
% \caption{Diagnostic comparison of marginal and group-adjusted residual library size dependence at the designed partition $(3{,}000,10{,}000,30{,}000)$ under library size-dependent diversity ($D=1$). Values are averaged over three simulated datasets for each setting. The group-adjusted criterion is used only to diagnose the behavior of the marginal relative-$R^2$ criterion and is not part of the proposed procedure.}
\label{tab:group_adjusted_Q_diagnostic}

\begin{tabular}{ccrr}
\toprule
$G$ & $p$
& Marginal $Q$
& Group-adjusted $Q$ \\
\midrule
0 & 0     & 0.208 & 0.206 \\
0 & 0.003 & 0.208 & 0.211 \\
0 & 0.300 & 0.029 & 0.210 \\
\midrule
1 & 0     & 0.207 & 0.206 \\
1 & 0.003 & 0.204 & 0.203 \\
1 & 0.300 & 0.037 & 0.183 \\
\bottomrule
\end{tabular}%

\end{table}

At small values of $p=0$ and $0.003$, the marginal and group-adjusted measures were nearly identical. In contrast, when $p=0.30$, the marginal value of $Q$ at the designed partition decreased sharply for both $G=0$ and $G=1$ cases. After adjusting for group, however, the corresponding values remained close to $0.18$--$0.21$ as $p$ increases. The decline in the marginal criterion at large $p$ does not indicate that the underlying library-size dependent structure has disappeared. Instead, the strong between-group difference increases the total variation in alpha diversity and thereby reduces the relative contribution attributed to library size on the marginal $R^2$ scale.

As an additional diagnostic, we repeated the fixed-$K$ and varying-$K$ optimization using the group-adjusted criterion. At $p=0.30$, both procedures selected $K=3$ in all diagnostic replications for both $G=0$ and $G=1$, with
small localization errors relative to the designed cutpoints. Recovery of the three-cutpoint structure under group adjustment supports the interpretation that the deterioration of the varying-$K$ procedure at large $p$ in the main analysis is driven primarily by the behavior of the marginal partial-$R^2$ criterion, rather than by disappearance of the depth structure or failure of the genetic algorithm. The group-adjusted criterion is used here only as a diagnostic and is not proposed as an alternative partition-selection rule.

\clearpage
\section{Sensitivity and Stability Analyses}

\subsection{Candidate grid sensitivity}

We examined sensitivity to the candidate rarefying depth grid using three grids containing 29, 55, and 109 depths, respectively. The analysis considered two representative settings with library size dependent underlying diversity and an association between group and library size: the null setting \(G=1,D=1,p=0\) and the alternative setting \(G=1,D=1,p=0.003\). The same simulated datasets were used across grid choices within each replicate, and all three grids contained the designed structural locations at 3,000, 10,000, and 30,000 reads.

For the primary fixed-\(K\) procedure, inferential conclusions were stable across the three grids. Under \(p=0.003\), rejection rates ranged from 0.280--0.300, 0.467--0.473, and 0.733--0.763 for \(N=200,400,\) and \(800\), respectively. Under \(p=0\), the corresponding ranges were 0.067--0.073, 0.060--0.067, and 0.093--0.107. Agreement in rejection decisions relative to the default grid ranged from 0.910 to 0.990. Although the optimized objective tended to decrease slightly as the grid became finer, downstream inferential conclusions changed little.

The varying-\(K\) procedure was more sensitive to the finest 109-point grid. In some settings, the attained objective was higher than under the default grid despite the larger search space, indicating increased optimization difficulty in the varying-\(K\) search. These results support the stability of the primary fixed-\(K\) procedure across reasonable candidate grids while reinforcing the greater computational sensitivity of the varying-\(K\) extension.

\subsection{GA initialization stability}

We further examined the stability of adaptive partition selection to stochastic optimization. For each of two representative settings with library size-dependent diversity (\(D=1\)) and group and library size association (\(G=1\)), corresponding to $p=0$ and $p=0.003$, respectively, we held the simulated dataset fixed and reran the optimization from 50 different genetic-algorithm initializations. Thus, any variation in the selected partition was attributable solely to stochastic optimization rather than Monte Carlo variation in the generated data.

The primary fixed-$K$ procedure was highly stable. For both settings, all 50 initializations selected the same three-cutpoint configuration, $(3000,8000,30000)$, yielding no variation in the optimized objective. The varying-$K$ procedure showed slightly higher sensitivity to initialization. Under $p=0$, $88\%$ of runs selected $K=3$, with 11 distinct cutpoint configurations across the 50 runs. Under $p=0.003$, all runs selected $K=3$, and the number of distinct configurations decreased to five. In both settings, the modal varying-$K$ configuration was $(3000,8000,30000)$. These results indicate that the primary fixed-$K$ implementation is effectively stable to genetic-algorithm initialization under the examined settings, whereas the more flexible varying-$K$ extension can exhibit modest optimization variability.

\begin{table}[!h]
\centering
\caption{Stability of adaptive partition selection across 50 genetic-algorithm initializations with the simulated dataset held fixed.}
\label{tab:ga_stability}
\begin{tabular}{clrrrr}
\toprule
$p$ & Method &
$\Pr(\hat K=3)$ &
\shortstack{Unique\\partitions} &
\shortstack{Modal partition\\frequency} &
\shortstack{Objective\\range} \\
\midrule
0
& Fixed-$K$
& 1.00 & 1  & 1.00 & 0.000 \\
0
& Varying-$K$
& 0.88 & 11 & 0.36 & 0.043 \\
0.003
& Fixed-$K$
& 1.00 & 1  & 1.00 & 0.000 \\
0.003
& Varying-$K$
& 1.00 & 5  & 0.58 & 0.026 \\
\bottomrule
\end{tabular}
\end{table}

\clearpage
\section{Supplementary Figures}

\begin{figure}[!h]
  \centering
  \includegraphics[width=\textwidth]{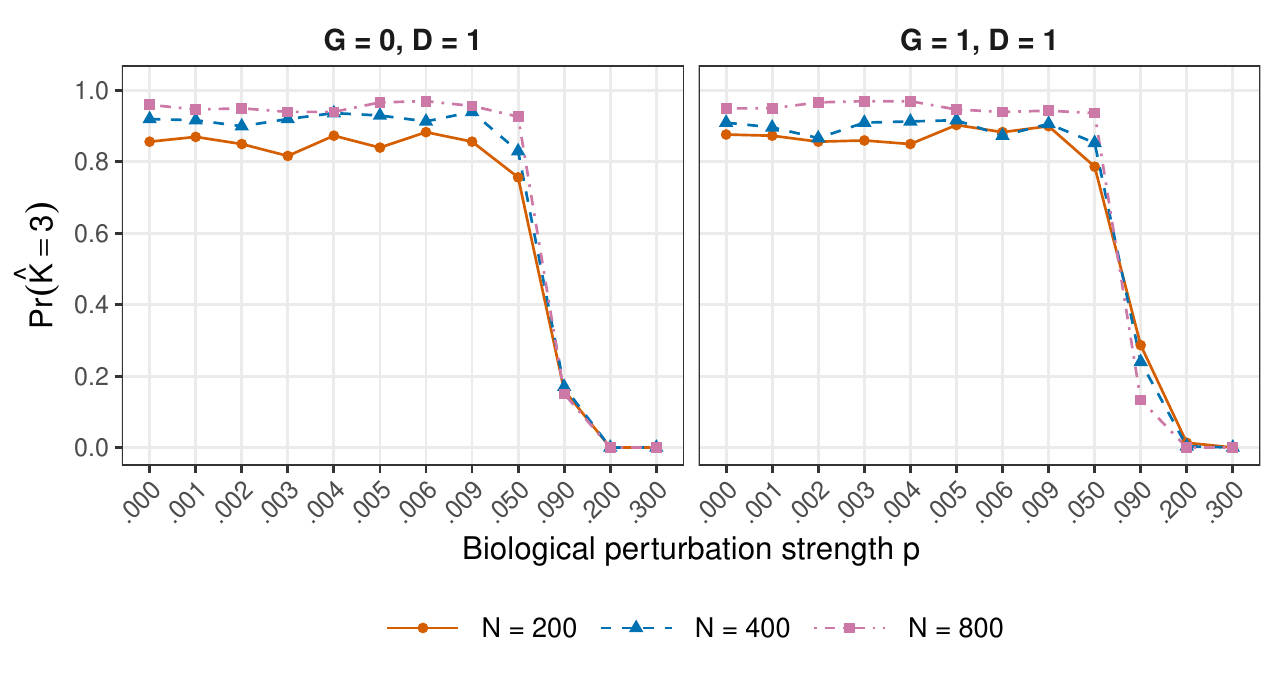}
  \caption{Recovery of the designed number of cutpoints under library size dependent diversity (\(D=1\)). Values show the proportion of replicates with \(\widehat K=3\).}
  \label{fig:structure_recovery_number}
\end{figure}

\clearpage

\begin{figure}[!h]
  \centering
  \includegraphics[width=\textwidth]{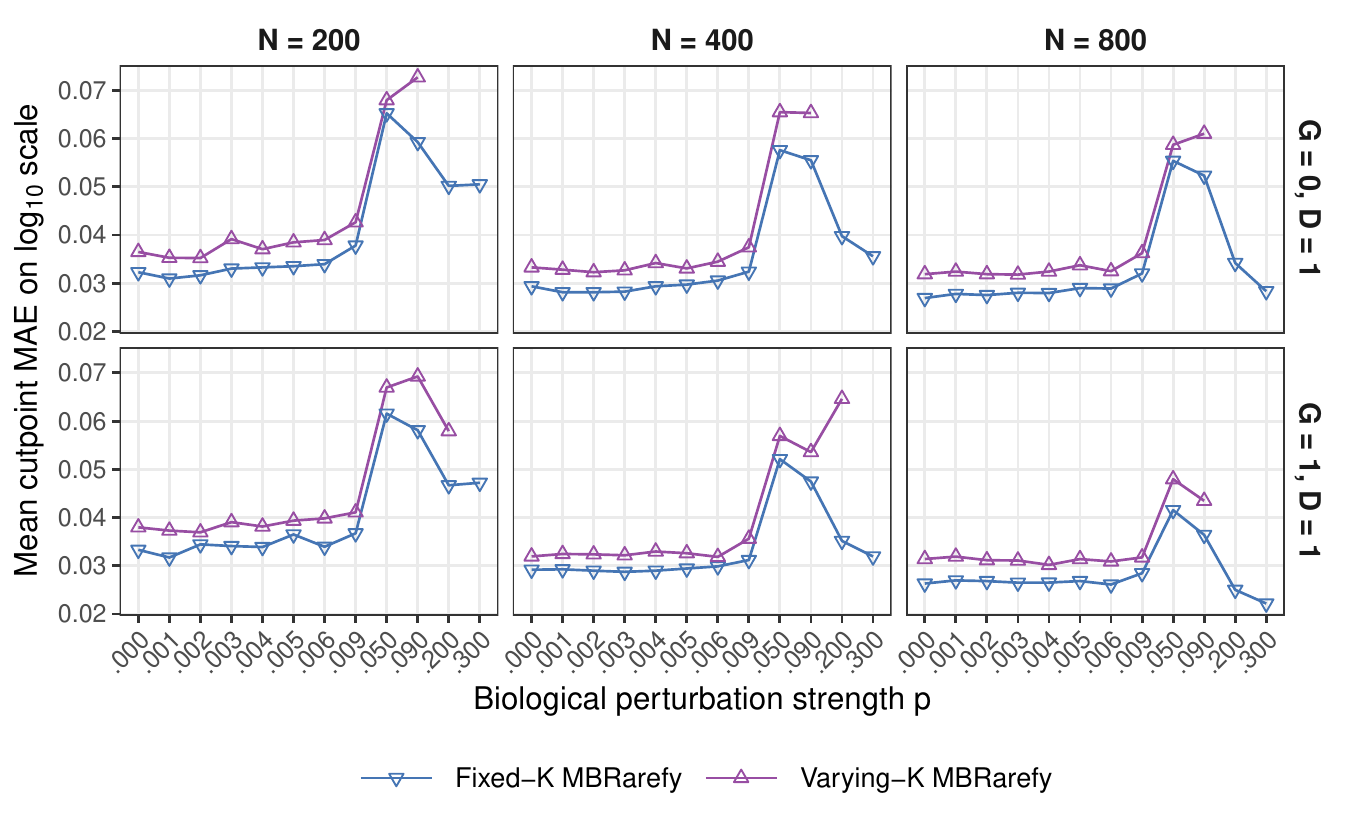}
    \caption{Recovery of the designed cutpoint locations under library-size-dependent diversity (\(D=1\)), measured by the mean absolute error on the \(\log_{10}\) depth scale. Smaller values indicate more accurate localization. For the varying-\(K\) procedure, the location error is computed only for replications in which three cutpoints were selected.}
  \label{fig:structure_recovery_location}
\end{figure}

\clearpage
\section{Other Supplementary Tables}

\begin{table*}[!ht]
\centering
\caption{Comparison of fixed and varying-$K$ adaptive partitions in the TCR application. All analyses used a minimum bin size of 10. $Q$ measures residual library-size dependence, $F$ is the fragmentation penalty, and  $J=Q+F$ is the partition-selection objective. For each diversity metric, the percentage change in $J$ is calculated relative to the corresponding fixed-$K=3$ solution.}
\label{tab:TCR_partition_objective}
\small

\resizebox{\textwidth}{!}{%
\begin{tabular}{llccccccc}
\toprule
Metric
& Procedure
& $K$
& Cutpoints ($\times10^6$ reads)
& Bin sizes
& $Q$
& $F$
& $J$
& $\Delta J$ (\%) \\
\midrule

Richness
& Fixed
& 3
& $2.0,\ 4.3,\ 7.2$
& $143,\ 181,\ 176,\ 163$
& 0.0215
& 0.1597
& 0.1811
& -- \\

Richness
& Fixed
& 5
& $1.9,\ 3.0,\ 4.5,\ 6.3,\ 8.1$
& $127,\ 105,\ 100,\ 111,\ 112,\ 108$
& 0.0189
& 0.2649
& 0.2838
& $+56.7$ \\

Richness
& Varying
& 1
& $3.2$
& $243,\ 420$
& 0.0485
& 0.0607
& 0.1092
& $-39.7$ \\

\midrule

Shannon
& Fixed
& 3
& $2.0,\ 4.5,\ 7.1$
& $143,\ 189,\ 163,\ 168$
& 0.0101
& 0.1600
& 0.1701
& -- \\

Shannon
& Fixed
& 5
& $1.9,\ 3.2,\ 4.8,\ 6.7,\ 8.4$
& $127,\ 116,\ 109,\ 117,\ 105,\ 89$
& 0.0101
& 0.2671
& 0.2772
& $+63.0$ \\

Shannon
& Varying
& 1
& $3.2$
& $243,\ 420$
& 0.0145
& 0.0607
& 0.0752
& $-55.8$ \\

\bottomrule
\end{tabular}
}
\end{table*}

\begin{table*}[!hb]
\centering
\caption{Sensitivity of the estimated CMV association to additional adjustment for original library size in the TCR application. The unadjusted estimates are those obtained under each normalization procedure in the primary analysis. The adjusted estimates additionally control for $\log_{10}$ original library size; for multi-bin procedures, bin-specific adjusted CMV coefficients were combined using inverse-variance weighting. Relative change is the absolute percentage change in the CMV coefficient after additional library-size adjustment. Richness coefficients are reported on the $\times 10^{-4}$ scale.}
\label{tab:TCR_CMV_sensitivity}
\small

\resizebox{\textwidth}{!}{%
\begin{tabular}{llrrr}
\toprule
Metric
& Method
& Unadjusted CMV estimate
& Adjusted CMV estimate
& Relative change (\%) \\
\midrule

\multirow{6}{*}{Richness ($\times10^{-4}$)}
& No rarefying
& $-1.146$
& $-2.359$
& 105.9 \\

& Overall rarefying
& $-1.950$
& $-2.260$
& 15.9 \\

& Prespecified multi-bin
& $-1.455$
& $-1.448$
& 0.5 \\

& Fixed-$K$ ($K=3$)
& $-0.852$
& $-0.915$
& 7.4 \\

& Fixed-$K$ ($K=5$)
& $-0.760$
& $-0.808$
& 6.3 \\

& Varying-$K$
& $-0.769$
& $-0.880$
& 14.4 \\

\midrule

\multirow{6}{*}{Shannon}
& No rarefying
& $-0.730$
& $-0.790$
& 8.1 \\

& Overall rarefying
& $-0.758$
& $-0.778$
& 2.7 \\

& Prespecified multi-bin
& $-0.736$
& $-0.742$
& 0.8 \\

& Fixed-$K$ ($K=3$)
& $-0.721$
& $-0.733$
& 1.6 \\

& Fixed-$K$ ($K=5$)
& $-0.734$
& $-0.749$
& 2.0 \\

& Varying-$K$
& $-0.733$
& $-0.745$
& 1.6 \\

\bottomrule
\end{tabular}
}
\end{table*}

\clearpage

\begin{table*}[!ht]
\centering
\caption{Comparison of fixed- and varying-$K$ adaptive partitions in the wild baboon gut microbiome application.
All analyses used a minimum bin size of 10.
$Q$ measures residual library-size dependence, $F$ is the fragmentation penalty, and
$J=Q+F$ is the partition-selection objective.
For each diversity metric, the percentage change in $J$ is calculated relative to the corresponding fixed-$K=3$ solution.}
\label{tab:baboon_partition_objective}
\small

\resizebox{\textwidth}{!}{%
\begin{tabular}{llccccccc}
\toprule
Metric
& Procedure
& $K$
& Cutpoints ($\times10^3$ reads)
& Bin sizes
& $Q$
& $F$
& $J$
& $\Delta J$ (\%) \\
\midrule

Richness
& Fixed
& 3
& $58,\ 79,\ 100$
& $99,\ 169,\ 146,\ 171$
& 0.0170
& 0.1684
& 0.1854
& -- \\

Richness
& Fixed
& 5
& $58,\ 69,\ 81,\ 96,\ 114$
& $99,\ 82,\ 102,\ 99,\ 102,\ 101$
& 0.0167
& 0.2650
& 0.2818
& $+52.0$ \\

Richness
& Varying
& 1
& $87$
& $331,\ 254$
& 0.0719
& 0.0545
& 0.1264
& $-31.8$ \\

\midrule

Shannon
& Fixed
& 3
& $62,\ 82,\ 107$
& $125,\ 171,\ 153,\ 136$
& 0.0133
& 0.1609
& 0.1741
& -- \\

Shannon
& Fixed
& 5
& $58,\ 71,\ 84,\ 101,\ 121$
& $99,\ 101,\ 113,\ 107,\ 91,\ 74$
& 0.0052
& 0.2691
& 0.2743
& $+57.5$ \\

Shannon
& Varying
& 1
& $84$
& $313,\ 272$
& 0.0252
& 0.0532
& 0.0784
& $-55.0$ \\

\bottomrule
\end{tabular}
}
\end{table*}